\documentclass[prd,aps,floats,twocolumn,showpacs,superscriptaddress,preprintnumbers,showkeys,floatfix]{revtex4-1}

\usepackage[latin1]{inputenc}                    
\usepackage{graphicx}                            
\usepackage{latexsym}                            
\usepackage{amsfonts}                            
\usepackage{amssymb}                             
\usepackage{amsmath}                             
\usepackage[mathscr]{eucal}                      
\usepackage{dcolumn}                             
\usepackage{theorem}                             

\usepackage{subfigure,feynmf}
\usepackage{multirow}
\usepackage{slashbox}
\usepackage{placeins}
\usepackage{morefloats}
\usepackage{epstopdf}
                           
\begin{document}

\preprint{
\vbox{
\hbox{ADP-13-04/T824}
}}

\title[Chiral expansion of moments of quark distributions]{SU(3) chiral perturbation theory expansion of moments of quark distributions}
\author{P.E.~Shanahan}\affiliation{ARC Centre of Excellence in Particle Physics at the Terascale and CSSM, School of Chemistry and Physics,
  University of Adelaide, Adelaide SA 5005, Australia}
\author{A.W.~Thomas}\affiliation{ARC Centre of Excellence in Particle Physics at the Terascale and CSSM, School of Chemistry and Physics,
  University of Adelaide, Adelaide SA 5005, Australia}
\author{R.D.~Young}\affiliation{ARC Centre of Excellence in Particle Physics at the Terascale and CSSM, School of Chemistry and Physics,
  University of Adelaide, Adelaide SA 5005, Australia}

\begin{abstract}
We present formulae for the chiral extrapolation of spin-dependent and spin-independent moments of quark distributions of octet baryons, including loop corrections and counterterms to leading non-analytic order. This analysis allows for isospin breaking, and may be used for the chiral extrapolation of both $(2+1)$- and $(1+1+1)$-flavor lattice QCD results. An example of such an application is given, with the extrapolation formulae applied, using the finite-range regularization scheme, to recent $(2+1)$-flavor QCDSF/UKQCD Collaboration lattice results for the first spin-independent and first two spin-dependent Mellin moments.
\end{abstract}

\pacs{12.38.Gc, 14.20.Dh, 12.39.Fe}

\keywords{Charge Symmetry Breaking, Parton Distribution, Lattice QCD, Chiral Symmetry, Extrapolation}

\maketitle

%
\section{Introduction}



Understanding hadron structure, in particular the partonic structure
of baryons, remains a significant challenge in nuclear physics.
Of particular importance to experimental programs, 
especially for the analysis of the scattering of ultra-high-energy 
cosmic ray particles 
or of fixed target and colliding hadron beam experiments, 
is a quantitative understanding of parton distribution functions (PDFs). 
In the infinite momentum frame these parameterize the likelihood of 
a particular parton carrying the Bjorken momentum fraction $x$ 
at a renormalization scale $\mu$.

PDFs have been well determined 
experimentally~\cite{Adloff:2003uh,Ball:2008by,Martin:2007bv,Lai:2010vv} 
and widely studied within 
models~\cite{Schreiber:1991qx,Diakonov:1996sr,Gamberg:2003ey,Cloet:2005pp,Cloet:2007em,Bacchetta:2008af,Lorce:2011dv}.
However, ultimately one wants to determine them directly from QCD itself and 
lattice field theory is currently the only quantitative tool 
available with this facility. 
While it is not possible to calculate PDFs directly on the lattice, 
use of the operator product expansion allows moments of PDFs, 
which represent averages over the momentum fraction $x$ carried by 
the parton, to be 
evaluated~\cite{Detmold:2002nf,Bali:2012av,Horsley:2010th,Cloet:2012db,
Gockeler:2004wp,Bratt:2010jn,Aoki:2010xg,Alexandrou:2011nr}.

In order to compare moments from lattice simulations, 
performed on a finite four-dimensional 
grid, with experimental determinations, several extrapolations 
must be performed. Both the continuum extrapolation as lattice 
spacing $a \rightarrow 0$, and finite volume effects which account for 
the finite extent of the lattice, must be considered.
As most lattice simulations are still performed at larger than physical 
quark masses, an extrapolation down in quark or pseudoscalar mass to 
the physical point is also necessary. 
That particular extrapolation is the focus of this work.

Naive linear extrapolation of lattice results for the first several moments of quark distributions to physical quark masses originally indicated a systematic discrepancy of more than 30$\%$ compared with experiment~\cite{Detmold:2001jb}. This was remedied somewhat by the use of chiral perturbation theory and the development of extrapolation formulae which incorporate the appropriate chiral physics~\cite{Detmold:2001jb,Detmold:2003tm}.
Following discussion of the consequences for flavor properties in Ref.~\cite{Thomas:2000nw}, chiral corrections to PDF moments in the nucleon were developed in Refs.~\cite{Chen:2001eg,Arndt:2001ye,Diehl:2006js,Dorati:2007bk,Chen:2001pva,Burkardt:2012hk,Detmold:2005pt,Chen:2001gra,Beane:2004rf}. These analyses include pion loops and octet and decuplet baryon intermediate states. Flavor symmetry breaking expansions about the SU(3) flavor-symmetric point were developed in Ref.~\cite{Cooke:2012xv}.

In this article we extend previous developments of chiral extrapolation formulae for quark distribution moments to allow for isospin breaking. We develop the formalism in general terms for all octet baryons, and consider all spin-independent and spin-dependent Mellin moments.

In section \ref{sec:Moments}, we define moments of quark distribution functions. Section~\ref{sec:Chiextrap} describes the derivation of chiral extrapolation formulae for these moments, and the results are summarized in section~\ref{subsec:fitfunctions}. Finally, we illustrate one use of this work by applying the results to the chiral extrapolation of recent lattice simulation results from the QCDSF/UKQCD Collaborations~\cite{Horsley:2010th,Cloet:2012db,Bietenholz:2011qq} in section~\ref{sec:latticefits}. 

\section{Moments of quark distribution functions}
\label{sec:Moments}


With $q^B_{\uparrow (\downarrow)}$ representing the number density of quarks of flavour $q$ whose spin is parallel (antiparallel) to the longitudinal spin direction of a baryon $B$, the spin-independent ($q^B(x)$) and spin-dependent ($\Delta q^B(x)$) quark distribution functions are defined as

\begin{align}
q^B(x) = & q^B_{\uparrow}(x) + q^B_{\downarrow}(x),\\
\Delta q^B (x) = & q^B _{\uparrow} (x) - q^B_{\downarrow}(x),
\end{align}
where $x$ is the fraction of the momentum of baryon $B$ carried by the quarks.

The ($n-1$)th spin-independent (SI) and $m$th spin-dependent (SD) Mellin moments of the quark distribution functions are defined as

\begin{align}
\langle x ^{n-1} \rangle^B_q = & \int^1_0 dx x^{n-1} ( q^B(x) + (-1)^n \overline{q}^B(x)),\\
\langle x ^{m} \rangle^B_{\Delta q} = & \int^1_0 dx x^{m} (\Delta q^B(x) + (-1)^m \Delta \overline{q}^B(x)).
\end{align}

The operator product expansion allows these moments to be related to the matrix elements of local twist-2 operators $\mathcal{O}$ by
%
\begin{align}
\nonumber
\langle B (\vec{p}) | [ \mathcal{O}&_q^{\{ \mu_1 \dotsm \mu_n \}} - \textrm{Tr} ] | B (\vec{p}) \rangle  \\ \label{eq:twistoprel1}
&= 2 \langle x^{n-1} \rangle^B_q [ p^{\{\mu_1}\dotsm p^{\mu_n \}} -\textrm{Tr}],\\[5pt] \nonumber
\langle B (\vec{p}) | [ \mathcal{O}&_{\Delta q}^{\left\{ \mu_0 \dotsm \mu_m \right\}} - \textrm{Tr} ] | B (\vec{p}) \rangle  \\ \label{eq:twistoprel2}
&=  2 \langle x^{m} \rangle^B_{\Delta q} M_B [S^{\{ \mu_0} p^{\mu_1}\dotsm p^{\mu_m \}} -\textrm{Tr}],
\end{align}
%
where $p^\mu$, $S^\mu$ and $M_B$ denote the momentum, spin and mass of the baryon $B$, the braces $\{ \ldots \}$ indicate total symmetrization of Lorentz indices, and trace terms involving $g^{\mu_i\mu_j}$ are subtracted to ensure that the operators transform irreducibly under the Lorentz group.
The twist-2 operators are defined as
\begin{align}
\label{eq:twistop1}
\mathcal{O}_q^{\mu_1 \dotsm \mu_n} & =i^{n-1} \overline{q} \gamma^{\mu_1} \overleftrightarrow{D}^{\mu_2} \dotsm \overleftrightarrow{D}^{\mu_n} q,\\ \label{eq:twistop2}
\mathcal{O}_{\Delta q}^{\mu_0 \dotsm \mu_m} & =i^{m} \overline{q} \gamma_5 \gamma^{\mu_0} \overleftrightarrow{D}^{\mu_1} \dotsm \overleftrightarrow{D}^{\mu_m} q,  
\end{align}
where $\overleftrightarrow{D}=\frac{1}{2} \left( \overrightarrow{D}-\overleftarrow{D} \right)$.

Hadronic matrix elements of these operators may be determined from lattice QCD using standard techniques. Given a suitable extrapolation to the physical point, such calculations give information about parton distributions directly from QCD itself.



\section{Chiral behaviour of quark distribution moments}
\label{sec:Chiextrap}

Here we outline the derivation of chiral extrapolation formulae for the quark distribution moments. This is done by first developing the extrapolation of the matrix elements of the relevant twist-2 operators shown in Eqs.~(\ref{eq:twistop1}) and (\ref{eq:twistop2}). We allow for isospin-breaking, that is, for $m_u \ne m_d$, so the results of this work may be applied to both $(2+1)$- and $(1+1+1)$-flavour lattice simulations of these moments. 

\subsection{Heavy baryon chiral perturbation theory}

To develop a chiral extrapolation of the parton distribution moments we include the twist-two operators given in Eqs.~(\ref{eq:twistop1}) and (\ref{eq:twistop2}) into the chiral Lagrange density of heavy-baryon chiral perturbation theory. This formalism, developed in Refs.~\cite{Jenkins:1990jv,Jenkins:1991es}, treats the baryons as heavy fields and has a consistent power counting expansion within which S-matrix elements can be expanded, below the symmetry-breaking scale $\Lambda_\chi$, in powers of derivatives and the quark mass matrix $m_q$.


We briefly review relevant details of the heavy-baryon formalism. The heavy-baryon chiral Lagrange density is written in terms of the (formally velocity-dependent) baryon fields
\begin{equation}
\textbf{B} = \left(
\begin{array}{ccc}
\frac{\Sigma^0}{\sqrt{2}} + \frac{\Lambda}{\sqrt{6}} & \Sigma^+ & p \\
\Sigma^- & \frac{\Lambda}{\sqrt{6}}-\frac{\Sigma^0}{\sqrt{2}} & n \\
\Xi^- & \Xi^0 & -\sqrt{\frac{2}{3}} \Lambda \\
\end{array}
\right)
\end{equation}
which may be expressed in tenor form as
\begin{equation}
B_{abc}=\frac{1}{\sqrt{6}}\left(\epsilon_{abd}\textbf{B}^d_{~c} + \epsilon_{acd} \textbf{B}^d_{~b} \right),
\end{equation}
and the pseudoscalar fields
\begin{equation}
\Phi = \frac{1}{\sqrt{2}}\left( 
\begin{array}{ccc}
\frac{\pi^0}{\sqrt{2}}+\frac{\eta}{\sqrt{6}} & \pi^+ & K^+ \\
\pi^- & \frac{\eta}{\sqrt{6}} - \frac{\pi^0}{\sqrt{2}} & K^0 \\
K^- & \overline{K}^0 & -\sqrt{\frac{2}{3}} \eta \\
\end{array}
\right),\\
\end{equation}
where
\begin{equation}
\Sigma =\textrm{exp} \left(\frac{2 i \Phi}{f} \right) = \xi^2. \\
\end{equation}
Under SU(3)$_L$ $\times$ SU(3)$_R$, the fields transform as
\begin{align}
\Sigma &\rightarrow L\Sigma R, \\
 B &\rightarrow U B U^\dagger, \\ \label{eq:U}
\xi &\rightarrow L \xi U^\dagger = U \xi R^\dagger,
\end{align}
with $U$ implicitly defined by Eq.~(\ref{eq:U}).

Decuplet baryons may be included by way of a Rarita-Schwinger field, represented by the tensor $T^{ijk}$: 
\begin{widetext}
\begin{equation}
T = \left\{
\left(
\begin{array}{ccc}
\Delta^{++} & \frac{\Delta^+}{\sqrt{3}} & \frac{\Sigma^{*+}}{\sqrt{3}} \\
\frac{\Delta^+}{\sqrt{3}} & \frac{\Delta^0}{\sqrt{3}} & \frac{\Sigma^{*0}}{\sqrt{6}} \\
\frac{\Sigma^{*+}}{\sqrt{3}} & \frac{\Sigma^{*0}}{\sqrt{6}} & \frac{\Xi^{*0}}{\sqrt{3}}\\
\end{array}
\right)
,
\left(
\begin{array}{ccc}
\frac{\Delta^+}{\sqrt{3}} & \frac{\Delta^0}{\sqrt{3}} & \frac{\Sigma^{*0}}{\sqrt{6}} \\
\frac{\Delta^0}{\sqrt{3}} & \Delta^- & \frac{\Sigma^{*-}}{\sqrt{3}} \\
\frac{\Sigma^{*0}}{\sqrt{6}} & \frac{\Sigma^{*-}}{\sqrt{3}} & \frac{\Xi^{*-}}{\sqrt{3}}\\
\end{array}
\right)
,
\left(
\begin{array}{ccc}
\frac{\Sigma^{*+}}{\sqrt{3}} & \frac{\Sigma^{*0}}{\sqrt{6}} & \frac{\Xi^{*0}}{\sqrt{3}} \\
\frac{\Sigma^{*0}}{\sqrt{6}} & \frac{\Sigma^{*-}}{\sqrt{3}} & \frac{\Xi^{*-}}{\sqrt{3}} \\
\frac{\Xi^{*0}}{\sqrt{3}} & \frac{\Xi^{*-}}{\sqrt{3}} & \Omega^- \\
\end{array}
\right)
\right\}.
\end{equation}
\end{widetext}
This field contains both spin$-1/2$ and spin$-3/2$ pieces; the spin$-1/2$ pieces are projected out by the constraint $\gamma_\mu T^\mu = 0$. Under SU(3)$_L$ $\times$ SU(3)$_R$, $T^\mu_{abc} \rightarrow U^d_aU^e_bU^f_cT^\mu_{def}$.

The interactions of the octet baryons, decuplet baryons and mesons are encoded in the following terms of the usual lowest-order effective Lagrangian~\cite{Jenkins:1990jv} (where we have retained only those terms needed for our calculation):
\begin{equation}
\label{eq:FD}
2D \textrm{Tr} \overline{\textbf{B}} S^\mu \{ \mathcal{A}_\mu, \textbf{B}\} + 2F \textrm{Tr} \overline{\textbf{B}}S^\mu [\mathcal{A}_\mu, \textbf{B}],
\end{equation}
\begin{equation}
\label{eq:C}
\sqrt{\frac{3}{2}} \mathcal{C} \left[ ( \overline{T}^\nu \mathcal{A}_\nu B ) +( \overline{B} \mathcal{A}_\nu T^\nu ) \right],
\end{equation}
where
\begin{equation}
\mathcal{A}_\mu = \frac{i}{2} \left( \xi \partial_\mu \xi^\dagger - \xi^\dagger \partial_\mu \xi \right)
\end{equation}
and flavor space contractions denoted by brackets $(\ldots)$ are given by
\begin{align}
\nonumber
(\overline{B} Y B ) &= \overline{B}^{kji}Y^l_i B_{ljk},\\
(\overline{B} B Y ) &= \overline{B}^{kji}Y^l_k B_{ijl},
\end{align}
where $B$ represents either the octet or decuplet baryon tensor.

The quark mass matrix $m_q$ is defined as
\begin{equation}
m_q= \left(
\begin{array}{ccc}
m_u & & \\
 & m_d & \\
& & m_s \\
\end{array}
\right),
\end{equation}
and
\begin{equation}
M= \frac{1}{2} \left( \xi m_q \xi + \xi^\dagger m_q \xi^\dagger \right).
\end{equation}
It should be noted that that
\begin{itemize}
\item $S^\mu$ is dimensionless, and $\overline{B}\gamma_5\gamma^\mu B=-2\overline{B} S^\mu B$. Note that this differs from the convention chosen in Ref.~\cite{Cloet:2012db}, where $S^2 \propto M_B^2$,
\item The baryon states are normalized such that $\overline{B} B \propto \textrm{ the baryon mass } M_B$,
\item Given the normalization for the pseudoscalar fields defined above, a chiral perturbation theory estimate of the pion decay constant in the chiral limit is $f_{\textrm{chiral}}=0.0871$~GeV~\cite{Amoros:2001cp}.
\end{itemize}

\subsection{Twist-two effective operators}
\label{sec:effops}

The twist-two operators $\mathcal{O}$, given in Eqs.~(\ref{eq:twistop1}) and (\ref{eq:twistop2}), must be represented within the framework of chiral effective field theory. That is, in the low energy effective theory, the quark bilinear operators are matched onto hadronic analogues constructed to obey the same symmetry transformation properties; under SU(3)$_L$ $\times$ SU(3)$_R$ the effective operators must transform as $(8,1) \oplus (1,8)$.

To describe each independent flavor operator, we define
\begin{equation}
\lambda^q = \frac{1}{2} \left(\xi \overline{\lambda}^q \xi^\dagger + \xi^\dagger\overline{\lambda}^q \xi \right),
\end{equation}
where for each quark flavor $q$, $\overline{\lambda}^q$ is given by
\begin{equation}
\overline{\lambda}^u = \left(\begin{array}{ccc} 1 & & \\ & & \\ & & \\ \end{array}\right)
\hspace{0.3cm}
\overline{\lambda}^d = \left(\begin{array}{ccc} & & \\ & 1 & \\ & & \\ \end{array}\right)
\hspace{0.3cm}
\overline{\lambda}^s = \left(\begin{array}{ccc} & & \\ & & \\ & & 1 \\ \end{array}\right).
\end{equation}
Effective operators corresponding to the isovector moment, for example, would have operator insertions containing $\lambda  = \lambda^u-\lambda^d$ ($=\lambda_3$ in the usual Gell-Mann basis for SU(3)).
 
It should be noted that the expressions given in the following sections differ from those of other works~\cite{Chen:2001gra,Chen:2001eg,Arndt:2001ye} by factors of the baryon mass $M_B$. We have chosen our convention so as to make dimensionless the unknown coefficients, $\alpha^{(n)}$, $\beta^{(n)}$, $\sigma^{(n)}$, $b_i^{(n)}$, which appear in the effective matrix elements.
%

\subsubsection{Spin-independent moments}

The terms listed in this section represent local operators that contribute to matrix elements of the trace-subtracted spin-independent twist-2 operators ($\mathcal{O}_q^{\{\mu_1 \ldots \mu_n\}}-\textrm{Tr}$). All terms involving zero or one mass insertion $M$ are included. The brackets $\{ \ldots \}$ representing total symmetrization of the enclosed Lorentz indices may also be written as `+ permutations' where this is notationally more convenient. This always indicates the symmetric sum with no normalization factor, i.e., $\{\mu \nu\}=\mu\nu + \nu\mu = (\mu\nu + \textrm{permutations})$. Superscripts $(n)$ on the undetermined coefficients indicate that these are distinct for each operator, that is, $\alpha^{(0)} \ne \alpha^{(1)}$ etc.

At leading order, the relevant effective operators contributing to the matrix elements are
\begin{align}
\nonumber
\left[ \vphantom{\sigma^{(n)}}\right.  &\alpha^{(n)} (\overline{B} B \lambda_q ) + \beta^{(n)} ( \overline{B} \lambda_q B)  \\ \label{eq:Bins}
& \left. {} + \sigma^{(n)} (\overline{B}B) \textrm{Tr} (\lambda_q)\right]  p^{\{\mu_1} \ldots p^{\mu_n\}} - \textrm{Tr},
\end{align}
the $\mathcal{O}(m_q)$ counterterms are given by
\begin{align}
\label{eq:BinsM}
\nonumber
\left( \vphantom{b_1^{(n)}} \right. & b_1^{(n)} \textrm{Tr} \left[ \overline{B} \left[ \left[ \lambda_q , B \right], M \right] \right] 
+ b_2^{(n)} \textrm{Tr} \left[ \overline{B} \{ \left[ \lambda_q , B \right], M \} \right]\\ \nonumber
&+ b_3^{(n)} \textrm{Tr} \left[ \overline{B} \left[ \{ \lambda_q , B \}, M \right] \right]
+ b_4^{(n)} \textrm{Tr} \left[ \overline{B} \{ \{ \lambda_q , B \}, M \} \right]\\ \nonumber
&+ b_5^{(n)} \textrm{Tr} \left[ \overline{B} B \right] \textrm{Tr} \left[ \lambda_q M \right]
+ b_6^{(n)} \textrm{Tr} \left[ \overline{B} B \lambda_q \right]\textrm{Tr} \left[ M \right]\\ \nonumber
&+ b_7^{(n)} \textrm{Tr} \left[ \overline{B} \lambda_q B \right]\textrm{Tr} \left[ M \right]
+ b_8^{(n)} \textrm{Tr} \left[ \overline{B} M B \right]\textrm{Tr} \left[ \lambda_q \right]\\ \nonumber
&+ b_9^{(n)} \textrm{Tr} \left[ \overline{B} B M \right]\textrm{Tr} \left[ \lambda_q \right]\\
&+b_{10}^{(n)} \textrm{Tr} \left[ \overline{B} \lambda_q \right]\textrm{Tr} \left[ M B \right]\left. \vphantom{b_1^{(n)}}\right) p^{\{\mu_1}\ldots p^{\mu_n\}} - \textrm{Tr},
\end{align}
and the decuplet insertions may be represented by
\begin{align}
\nonumber
&\gamma^{(n)}\left(\overline{T}^\nu \lambda_q T_\nu \right)p^{\{\mu_1}\ldots p^{\mu_n\}} \\ \label{eq:sidec}
& {}+ \gamma'^{(n)}M_B^2\left(\overline{T}^{\{\mu_1}\lambda_q T^{\mu_2}\right)p^{\mu_3} \ldots p^{\mu_n\}} - \textrm{Tr}.
\end{align}
Clearly, because of the number of available indices, $\gamma'^{(1,2)}=0$.

\subsubsection{Spin-dependent moments}

The spin-dependent operators have effective matrix elements which have a very similar structure to those given in the previous section for the spin-independent case. The term analogous to that of Eq.~(\ref{eq:Bins}) has the form
\begin{align}
\nonumber
\left[ \vphantom{\alpha^{(m)}} \right.& \Delta\alpha^{(m)} ( \overline{B} S^{\mu_0} B \lambda_q ) + \Delta\beta^{(m)} ( \overline{B} S^{\mu_0} \lambda_q B) \\ \nonumber
& + \ \Delta\sigma^{(m)} (\overline{B} S^{\mu_0} B) \textrm{Tr} (\lambda_q)\left.\vphantom{\alpha^{(m)}} \right] p^{\mu_1}\ldots p^{\mu_m} \\
& \! \! \!+ \textrm{permutations} - \textrm{Tr}.
\end{align}
For $m=0$, we note that by the Goldberger-Treiman relation the zeroth moments of the spin-dependent moments are related to the meson-baryon coupling constants by
\begin{align}
\Delta \alpha^{(0)}= & 2\left(\frac{2}{3}D+2F\right), \\
\Delta \beta^{(0)}= & 2\left(-\frac{5}{3}D+F\right), \\
\end{align}
where $F$ and $D$ are defined by Eq.~(\ref{eq:FD}).

The form of the effective operator matrix elements with insertions of the quark mass matrix $M$ is again entirely analogous to that for the spin-independent case:
\begin{widetext}
\begin{align}
\nonumber
& \left(\Delta b_1^{(m)} \right. \textrm{Tr} \left[ \overline{B} S^{\mu_0} \left[ \left[ \lambda_q , B \right], M \right] \right] 
+ \Delta b_2^{(m)} \textrm{Tr} \left[ \overline{B} S^{\mu_0} \{ \left[ \lambda_q , B \right], M \} \right]
+ \Delta b_3^{(m)} \textrm{Tr} \left[ \overline{B} S^{\mu_0} \left[ \{ \lambda_q , B \}, M \right] \right]\\ \nonumber
&+ \Delta b_4^{(m)} \textrm{Tr} \left[ \overline{B} S^{\mu_0} \{ \{ \lambda_q , B \}, M \} \right]
+ \Delta b_5^{(m)} \textrm{Tr} \left[ \overline{B} S^{\mu_0} B \right] \textrm{Tr} \left[ \lambda_q M \right]
+ \Delta b_6^{(m)} \textrm{Tr} \left[ \overline{B} S^{\mu_0} B \lambda_q \right]\textrm{Tr} \left[ M \right]\\ \nonumber
&+ \Delta b_7^{(m)} \textrm{Tr} \left[ \overline{B} S^{\mu_0} \lambda_q B \right]\textrm{Tr} \left[ M \right]
+ \Delta b_8^{(m)} \textrm{Tr} \left[ \overline{B} S^{\mu_0} M B \right]\textrm{Tr} \left[ \lambda_q \right]
+ \Delta b_9^{(m)} \textrm{Tr} \left[ \overline{B} S^{\mu_0} B M \right]\textrm{Tr} \left[ \lambda_q \right]\\ 
&+\Delta \left. b_{10}^{(m)} \textrm{Tr} \left[ \overline{B} S^{\mu_0} \lambda_q \right]\textrm{Tr} \left[ M B \right]\right) p^{\mu_1}\ldots p^{\mu_m} + \textrm{permutations} - \textrm{Tr}. 
\end{align}
\end{widetext}

Decuplet contributions may be represented by
\begin{align}
\nonumber
& \Delta\gamma^{(m)} \left(\overline{T}^\nu S^{\{\mu_0} \lambda_q T_\nu \right)p^{\mu_1} \ldots p^{\mu_m\}} \\ \label{eq:sddec}
& {}+ \Delta \gamma'^{(m)}M_B^2\left(\overline{T}^{\{\mu_1}S^{\mu_0}\lambda_q T^{\mu_2}\right)p^{\mu_3} \ldots p^{\mu_m\}} - \textrm{Tr}.
\end{align}
Clearly, because of the number of available indices, $\Delta \gamma'^{(0,1)}=0$. Other approximate relations between the unknown coefficients may be derived using SU(6) symmetry. In our numerical calculations, for example, we set $\Delta\gamma^{(0)}=2\mathcal{H}=-6D$. The analogous relation for the first moment is $\Delta\gamma^{(1)}=-\frac{3}{2}(\Delta\alpha^{(1)}-2\Delta\beta^{(1)})$.

Transitions between octet and decuplet baryons via an operator insertion are also allowed in the spin-dependent case, and are represented by the effective matrix element
\begin{align}
\nonumber
&\sqrt{\frac{3}{2}} \omega^{(m)}\left[ ( \overline{T}^{\mu_0} \lambda_q B ) +( \overline{B} \lambda_q T^{\mu_0} ) \right]p^{\mu_1}\ldots p^{\mu_m} \\
&  + \textrm{permutations} - \textrm{Tr}.
\end{align}
 Here $\omega^{(0)}=\mathcal{C}$ is the same parameter which appeared in Eq.~(\ref{eq:C}). For our numerical results we use the SU(6) approximation, setting $\omega^{(1)}=-\frac{1}{2}(\Delta\alpha^{(1)}-2\Delta\beta^{(1)})$. 

\begin{widetext}

\unitlength=1mm
\begin{figure}[tbf]
\centering
\subfigure[]{
\includegraphics{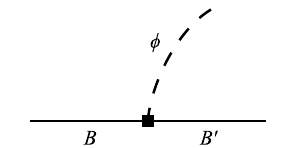}
\label{fig:vB}
}
\subfigure[]{
\includegraphics{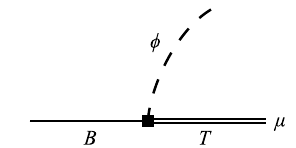}
\label{fig:vT}
}
\subfigure[]{
\includegraphics{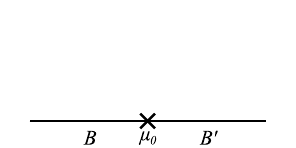}
\label{fig:cB}
}
\subfigure[]{
\includegraphics{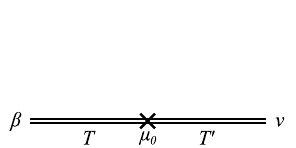}
\label{fig:cT}
}
\subfigure[]{
\includegraphics{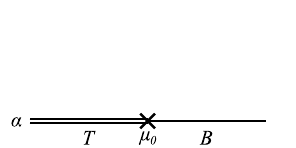}
\label{fig:cBT}
}
\subfigure[]{
\includegraphics{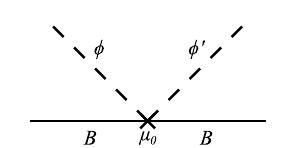}
\label{fig:phi}
}
\caption{Vertices and operator insertions which appear in the leading non-analytic contributions to moments of quark distribution functions. Solid squares indicate leading-order strong interaction vertices, while the cross indicates an insertion of the twist-two operator. This insertion carries a Lorentz index $\mu_0$ in the spin-dependent case only.}
\label{fig:vertices}
\end{figure}
\end{widetext}

\subsection{Feynman rules}
\label{subsec:feynrules}

Feynman rules relevant to the chiral extrapolation of matrix elements of twist-2 operators may be read directly from the effective operator matrix element terms given in section~\ref{sec:effops}.

In standard heavy baryon chiral perturbation theory the baryon propagators and baryon-meson vertices are given by:
\begin{align}
\label{eq:feyn1}
\nonumber
& \textrm{Octet Propagator:}  &\frac{i}{k\cdot v+ i \epsilon} \\ \nonumber
& \textrm{Decuplet Propagator:}  &\frac{iP^{\mu\nu}}{k \cdot v + \delta + i \epsilon} \\ \nonumber
& \textrm{Meson Propagator:} &\frac{i}{k^2-m_\phi^2 + i \epsilon}\\ \nonumber
& \textrm{$BB'\phi$ Vertex \ref{fig:vB}:} &\frac{k \cdot \mathcal{S}}{f}C_{BB'\phi}\\
& \textrm{$BT\phi$ Vertex \ref{fig:vT}:} & \frac{k_\mu}{f}C_{BT\phi}
\end{align}
where $v$ denotes the four-velocity of the heavy baryon $B$, $k^\mu$ refers to the momentum of the baryon or meson where the meson is outgoing from vertices, and $P^{\mu\nu}= (v^\mu v^\nu - g^{\mu \nu}) - \frac{4}{3}S^\mu S^\nu$ is a polarization projector. The labels \ref{fig:vB}, \ref{fig:vT} refer to the corresponding figures. We note that the flavour algebra is encompassed in the definitions of the (Clebsch-Gordan) coefficients $C$ which are given explicitly in appendix~\ref{app:coeffs}. Subscripts $B$, $T$ and $\phi$ on these coefficients label the octet baryon, decuplet baryon, and meson which appear in the corresponding vertex, while a subscript $O_q$ indicates that the coupling corresponds to an operator insertion.

The terms corresponding to operator insertion vertices differ for the spin-independent (SI) and spin-dependent (SD) cases. For the spin-independent operators
\begin{widetext}
\begin{align}
\label{eq:feyn2}
\nonumber
& \textrm{$BB\phi\phi_{\textrm{SI}}$ Vertex Insertion \ref{fig:phi}:}  &  \frac{1}{M_Bf^2}C_{BB\phi\phi O_q}^{(n)}p^{\{ \mu_1} \ldots p^{\mu_n\}} \\ \nonumber
& \textrm{$BB'_{\textrm{SI}}$ Operator Insertion \ref{fig:cB}:}  &  \frac{1}{M_B}C_{BB'O_q}^{(n)}p^{\{ \mu_1} \ldots p^{\mu_n\}}\\ \nonumber
& \textrm{$TT'_{\textrm{SI}}$ Operator Insertion \ref{fig:cT} \#1:}  &   \frac{1}{M_B}C_{TT'O_q}^{(n)}g_{\nu\beta} p^{\{ \mu_1} \ldots p^{\mu_n\}} \\ \nonumber
& \textrm{$TT'_{\textrm{SI}}$ Operator Insertion \ref{fig:cT} \#2:}  &  \frac{1}{M_B}C_{TT'O_q}^{(n)}g_\nu^{~\{\mu_1}g_\beta^{~\mu_2}p^{\mu_3} \ldots p^{\mu_n\}}. \\
\end{align}
Similarly, for the spin-dependent operators
\begin{align}
\label{eq:feyn3}
\nonumber
& \textrm{$BB\phi\phi_{\textrm{SD}}$ Vertex Insertion \ref{fig:phi}:} &  \frac{1}{f^2}C_{BB\phi\phi O_{\Delta q}}^{(m)} S^{\{\mu_0} p^{ \mu_1} \ldots p^{\mu_m\}} \\ \nonumber
& \textrm{$BB'_{\textrm{SD}}$ Operator Insertion \ref{fig:cB}:}  &  C_{BB'O_{\Delta q}}^{(m)}S^{\{\mu_0} p^{ \mu_1} \ldots p^{\mu_m\}} \\ \nonumber
& \textrm{$TT'_{\textrm{SD}}$ Operator Insertion \ref{fig:cT} \#1:} & C_{TT'O_{\Delta q}}^{(m)}g_{\nu\beta}S^{\{\mu_0} p^{ \mu_1} \ldots p^{\mu_m\}}  \\ \nonumber
& \textrm{$TT'_{\textrm{SD}}$ Operator Insertion \ref{fig:cT} \#2:} & C_{TT'O_{\Delta q}}^{(m)}g_\nu^{~\{\mu_1}g_\beta^{~\mu_2}S^{\mu_0} p^{ \mu_3} \ldots p^{\mu_m\}} \\
& \textrm{$TB_{\textrm{SD}}$ Operator Insertion \ref{fig:cBT}:}  & C_{TBO_{\Delta q}}^{(m)}g^{~\{\mu_0}_\alpha p^{ \mu_1} \ldots p^{\mu_m\}} .
\end{align}
\end{widetext}

The $TT'$ operator insertions labelled \#1 and \#2 correspond to the first and second terms of the decuplet effective operator contributions, respectively (see Eqs.~(\ref{eq:sidec}) and (\ref{eq:sddec})).

\subsection{Feynman Diagrams}
\label{subsec:feyndiag}

This section details the loop contributions, illustrated in Fig.~\ref{fig:diagrams2}, which are included in this calculation. Amongst these are loops with both octet and decuplet intermediate states, tadpole loops, and wavefunction renormalization terms. Diagrams~\ref{fig:mesloop}--\ref{fig:tads} contribute only to the odd$-n$ spin independent moments at order $m_\pi^{n+1}\textrm{log}(m_\pi)$, and are thus included only for the $n=1$ spin-independent moment. For this moment they serve to cancel the contributions of diagrams~\ref{fig:Bins}--\ref{fig:Tnormphi} to give the quark flavor sum rule. 

\begin{widetext}

\unitlength=1mm
\begin{figure}
\centering
\subfigure[]{
\includegraphics{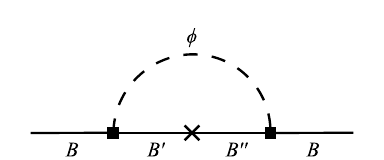}
\label{fig:Bins}
}
\subfigure[]{
\includegraphics{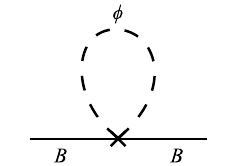}
\label{fig:tad}
}
\subfigure[]{
\includegraphics{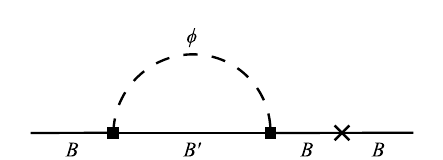}
\label{fig:Bnorm}
}
\subfigure[]{
\includegraphics{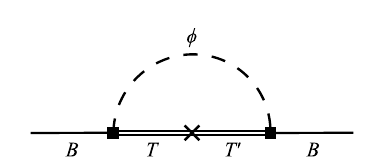}
\label{fig:Tins}
}
\subfigure[]{
\includegraphics{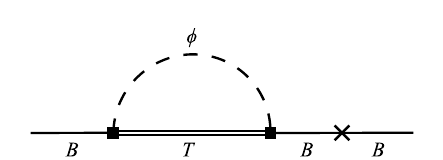}
\label{fig:Tnormphi}
}
\subfigure[]{
\includegraphics{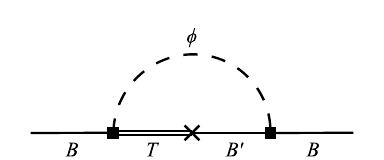}
\label{fig:change}
}
\subfigure[]{
\includegraphics{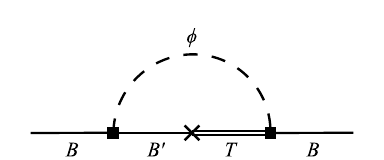}
\label{fig:change2}
}
\subfigure[]{
\includegraphics{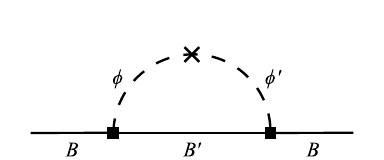}
\label{fig:mesloop}
}
\subfigure[]{
\includegraphics{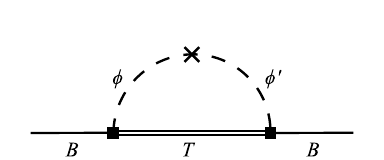}
\label{fig:mesloop2}
}
\subfigure[]{
\includegraphics{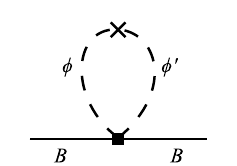}
\label{fig:tads}
}
\caption{Chiral loops included in the present calculation. Diagram~\ref{fig:Bins} is hereafter referred to as the `octet loop' diagram, Fig.~\ref{fig:Tins} is the `decuplet loop', and diagram~\ref{fig:tad} is referred to as the `tadpole' diagram. Diagrams~\ref{fig:Bnorm} and \ref{fig:Tnormphi} correspond to wavefunction renormalization. The transition diagrams, shown in Figures~\ref{fig:change} and \ref{fig:change2}, contribute only in the spin-dependent case. Diagrams~\ref{fig:mesloop}--\ref{fig:tads} are included for the $n=1$ spin-independent moment only, as explained in the text.}
\label{fig:diagrams2}
\end{figure}

\end{widetext}

\subsection{Loop integrals}
\label{subsec:loopint}

This section summarizes common integral expressions needed for the evaluation of diagrams included in our calculation.
Within the framework of finite-range regularization (FRR), we introduce a mass scale $\Lambda$ through a regulator $u(k)$ inserted into each integral expression~\cite{Thomas:2002sj,Stuckey:1996qr,Donoghue:1998bs,Leinweber:1998ej,Young:2002ib}. This regulator may take monopole, dipole, Gaussian, or sharp cutoff forms, for example. The parameter $\Lambda$ is related to the scale beyond which a formal expansion in powers of the Goldstone boson mass breaks down. Changing to dimensionally regularized integral expressions requires a simple substitution; details are given in~\cite{Borasoy:2002jv}.

Loops with octet baryon intermediate states involve the term
%


\begin{equation}
\label{eq:Jeq}
\int \frac{d^4k}{(2\pi)^4} \frac{k^i k^j}{(k_0-i\epsilon)^2 (k^2-m_\phi^2+i \epsilon)} \underset{FRR}{=} -i \delta^{ij} \frac{J(m^2)}{16 \pi^2},
\end{equation}
where
\begin{equation}
J(m^2) = \frac{4}{3} \int_0^\infty dk \frac{k^4u^2(k)}{(\sqrt{k^2+m^2})^3}
\end{equation}
with the finite-range regulator $u(k)$ inserted.  The normalization of $J(m^2)$ has been defined so that the non-analytic part is simply related to the common form of dimensionally regularized results, as $J(m^2) \underset{DR}{\rightarrow} m^2 \textrm{ln}(m^2/\mu^2)$.

Clearly, entirely analogous expressions can be written for integrals with decuplet propagators replacing one or more of the octet propagators in the above loop. We define
\begin{widetext}
\begin{align}
\int \frac{d^4k}{(2\pi)^4} \frac{k^i k^j}{(k_0+\delta-i\epsilon)(k_0-i\epsilon) (k^2-m_\phi^2+i \epsilon)} &\underset{FRR}{=} -i \delta^{ij} \frac{J_1(m^2,\delta)}{16 \pi^2}\\
\int \frac{d^4k}{(2\pi)^4} \frac{k^i k^j}{(k_0+\delta-i\epsilon)^2 (k^2-m_\phi^2+i \epsilon)} &\underset{FRR}{=} -i \delta^{ij} \frac{J_2(m^2,\delta)}{16 \pi^2}
\end{align}
\end{widetext}
\clearpage
where
\begin{align}
J_1(m^2,\delta) & = \frac{4}{3} \int_0^\infty dk \frac{k^4u^2(k)}{(\sqrt{k^2+m^2})^2(\sqrt{k^2+m^2}+\delta)}\\
J_2(m^2,\delta) & = \frac{4}{3} \int_0^\infty dk \frac{k^4u^2(k)}{(\sqrt{k^2+m^2})(\sqrt{k^2+m^2}+\delta)^2}
\end{align}
with one and two decuplet propagators respectively.

%
%
%


We also define
\begin{equation}
\label{eq:tadeq}
J_T(m^2)=4 \int_0^\infty dk \frac{k^2u^2(k)}{\sqrt{k^2+m^2}}
\end{equation}
which has the same non-analytic structure as $J$, i.e., $J_T(m^2) \underset{DR}{\rightarrow} m^2 \textrm{ln}(m^2/\mu^2)$. This integral will appear in the evaluation of tadpole loops in section~\ref{sec:tad}.

To make comparison with DR expressions clear, the integral replacement
\begin{equation}
\mathcal{I}(m_\phi) \rightarrow \widetilde{\mathcal{I}}(m_\phi)=\left[\mathcal{I}(m_\phi)-d_0^\Lambda-d_2^\Lambda m_\phi^2\right]
\end{equation}
is made, where $d_0^\Lambda$ and $d_2^\Lambda$ denote the leading analytic parts of the Taylor expansion of the integral, and $\mathcal{I}$ represents any of the integrals in Eqs.~(\ref{eq:Jeq})--(\ref{eq:tadeq}).
All expressions in this article should be taken to use the subtracted integral form. 
This renormalization process is described in detail for the case of baryon mass expansions in Ref.~\cite{Young:2002ib}. After the subtractions have been performed, the residual dependence of the chiral expansion on the FRR cutoff $\Lambda$ appears as inverse powers of $\Lambda$. This dependence may be minimized by fitting $\Lambda$ to lattice data to optimally reproduce the non-analytic structure displayed by the data. It may further be accounted for by allowing some variation in $\Lambda$, and by considering a range of regulator forms $u(k)$ which give different $\Lambda$-dependences.

We note that, by removing the unphysical short-distance part of loop diagrams, FRR has been shown to improve the convergence of the (traditionally poorly convergent) SU(3) chiral series~\cite{Donoghue:1998bs}, and consistently provides robust fits to lattice data at next-to-leading order. Nevertheless, one could check the size of next-to-next-to-leading order corrections to confirm that these contributions are small as expected.

\subsection{Loop contributions}

This section gives expressions for the contribution from each loop diagram shown in section~\ref{subsec:feyndiag}. Each term may be derived using the Feynman rules of section~\ref{subsec:feynrules}, and is written in terms of the subtracted integrals defined in section~\ref{subsec:loopint}. In each case, the subscripts $P$ and $U$ indicate the polarized (spin-dependent) and unpolarized cases, and the superscripts $8$ and $10$ indicate diagrams with octet and decuplet baryon intermediate states. All Clebsch-Gordon coefficients $C$, the momenta $p^{\mu_1} \ldots p^{\mu_{n/m}}$, and the associated symmetrization of Lorentz indices are omitted here.

\subsubsection{Wavefunction renormalization}

The contributions from wavefunction renormalization correspond to Figures~\ref{fig:Bnorm} and \ref{fig:Tnormphi}
\begin{align}
Z_{2,\{P,U\}}^8 & = \frac{1}{16\pi^2 f^2} \left( \frac{3}{8} \right)\widetilde{J}(m^2),\\
Z_{2,\{P,U\}}^{10} & = \frac{1}{16 \pi^2 f^2} \widetilde{J}_2(m^2,\delta).
\end{align}

\subsubsection{Tadpole loops}
\label{sec:tad}

The tadpole loop contributions correspond to Fig.~\ref{fig:tad}
\begin{align}
Z_{1,\{P,U\}}^{\textrm{tad}} &= \frac{1}{16 \pi^2 f^2} \left( \frac{1}{2} \right) \widetilde{J}_T(m^2).
\end{align}

\subsubsection{Octet intermediate state loops}

The contribution from Fig.~\ref{fig:Bins}, with an operator insertion into an octet baryon intermediate state, differs from the octet loop wavefunction renormalization term only in the spinor algebra.
\begin{align}
Z_{1,P}^{(8,8)}= &\frac{1}{16\pi^2 f^2} \left( -\frac{1}{8} \right)\widetilde{J}(m^2), \\
Z_{1,U}^{(8,8)}= &\frac{1}{16\pi^2 f^2} \left( \frac{3}{8} \right)\widetilde{J}(m^2).
\end{align}

\subsubsection{Decuplet intermediate state loops}

The contribution from decuplet loops with one operator insertion (Fig.~\ref{fig:Tins}) mimics that of the decuplet loop wavefunction renormalization term. We note that there is an extra $P^{\mu \nu}$ polarization projector in the spin algebra here, as there are two decuplet propagators (as opposed to the wavefunction renormalization term, which has one, but has the identical integral form $J_2$ because of the derivative with respect to external momentum). 
There are two separate terms which contribute to the decuplet loop (Fig.~\ref{fig:Tins}), arising from the two terms in each of Eqs.~(\ref{eq:sidec}) and~(\ref{eq:sddec}). Just as was done in labelling the Feynman rules in Eqs.~(\ref{eq:feyn2}) and~(\ref{eq:feyn3}), we label the two contributions as `1' and `2'.

\begin{align}
Z_{1,P1}^{(10,10)} = &\frac{1}{16 \pi^2 f^2} \left(-\frac{5}{9}\right) \widetilde{J}_2(m^2,\delta), \\
Z_{1,P2}^{(10,10)} = &\frac{1}{16 \pi^2 f^2} \left(\frac{1}{9}\right)\widetilde{J}_2(m^2,\delta) .
\end{align}
\begin{align}
Z_{1,U1}^{(10,10)} = &\frac{1}{16 \pi^2 f^2} \left(-1\right) \widetilde{J}_2(m^2,\delta), \\
Z_{1,U2}^{(10,10)}= &\frac{1}{16 \pi^2 f^2}  \left(\frac{1}{3}\right) \widetilde{J}_2(m^2,\delta).
\end{align}

\subsubsection{Octet-decuplet transition loops}

By symmetry, the contributions from diagrams \ref{fig:change} and \ref{fig:change2} are the same. These diagrams do not contribute in the spin-independent case.
\begin{equation}
\nonumber
Z_{1,P}^{(10,8)} =Z_{1,P}^{(8,10)} = \frac{1}{16 \pi^2 f^2} \left(\frac{2}{3}\right) \widetilde{J}_1(m^2,\delta).
\end{equation}

\subsection{Isospin breaking}
\label{subsec:isobreak}

In its most general form, after including a non-zero light quark mass splitting, $m_u \ne m_d$, the chiral perturbation theory expansion developed in this work will have separate couplings and integrals for each of the mesons $\pi^\pm,\pi^0,K^\pm,K^0,\eta$ in the mass-eigenstate basis. The $\pi^\pm$ and $K^\pm$ remain pairwise mass-degenerate. We recall that because of the necessary redefinition of the meson fields to remove $\pi^0-\eta$ mixing, the baryon-meson couplings will also receive contributions depending on the $\pi^0-\eta$ mixing angle $\epsilon$. Setting $\epsilon \rightarrow 0$ in all expressions will of course return the isospin-averaged results. Here we make explicit the dependence of $m_{\pi^0}$ and $m_\eta$ on the mixing angle $\epsilon$.

Consider the usual definition of the meson Lagrangian:
\begin{equation}
\label{mesonlagrangian}
\mathcal{L}=\frac{f^2}{8}\textrm{Tr}(\partial^\mu \Sigma^\dagger \partial_\mu \Sigma) + \lambda \mathrm{Tr}(m_q( \Sigma^\dagger + \Sigma)).
\end{equation}
Expanding this Lagrangian in powers of the meson field, the mass term can be written as 
\begin{align}
\mathcal{L}_{\textrm{mass}}= &B\textrm{Tr}(m_q \Phi^2)\\ \nonumber
 = & B(m_u+m_d)\pi^+ \pi^- + B(m_s+m_d)K^0 \overline{K}^0 \\ \nonumber
 & + B(m_s+m_u)K^+ K^- + \frac{B}{2}(m_u+m_d)(\pi^0)^2 \\ \nonumber
& +\frac{B}{6}(m_d+m_u+4m_s)\eta^2 \\
& +\frac{B}{\sqrt{3}}(m_u-m_d)\eta \pi^0,
\end{align}
where the final term indicates mixing between the $\pi^0$ and $\eta$ fields when $m_u \ne m_d$.

To identify the meson masses one must remove this mixing and bring the kinetic term into the canonical form via a field rotation
\begin{align}
\pi^0 & \rightarrow \pi^0 \textrm{cos} \epsilon - \eta \textrm{sin} \epsilon,\\
\eta & \rightarrow \pi^0 \textrm{sin} \epsilon + \eta \textrm{cos} \epsilon,
\end{align}
where the mixing angle $\epsilon$ is given by
\begin{equation}
\textrm{tan} 2 \epsilon = \frac{\sqrt{3}~(m_d-m_u)}{2m_s-(m_d+m_u)}.
\end{equation}

After performing this rotation, the SU(3) meson masses take the form:
\label{mesonmasses}
\begin{align}
m_{\pi^\pm}^2 = & B(m_u + m_d)\\ \nonumber
m_{\pi^0}^2  =& B(m_u+m_d) \\
&-\frac{2B}{3}(2m_s-(m_u+m_d))\frac{\textrm{sin}^2\epsilon}{\textrm{cos} 2 \epsilon}\\
m_{K^\pm}^2  =& B(m_s+m_u)\\
m_{K^0}^2 = & B(m_s+m_d)\\ \nonumber
m_{\eta}^2 = & \frac{B}{3}(4m_s+m_u+m_d)\\
& +\frac{2B}{3}(2m_s-(m_u+m_d))\frac{\textrm{sin}^2\epsilon}{\textrm{cos} 2 \epsilon},
\end{align}
where $m_{\pi^0}$ and $m_\eta$ now contain some dependence on the mixing angle $\epsilon$.

\subsection{Fit functions}
\label{subsec:fitfunctions}

In this section we give expressions for the chiral extrapolation of quark distribution moments.
The Clebsch-Gordon coefficients $C$ below are those given in the Feynman rules in Eqs.~(\ref{eq:feyn1}), (\ref{eq:feyn2}) and (\ref{eq:feyn3}). We emphasize that these coefficients are distinct for each moment, and are listed in appendix~\ref{app:coeffs}. In the expressions below, summation over repeated indices, e.g., $B'$, $T$, $\phi$ (but not $B$) is implied. The overall factor of 2 arises from the corresponding factor in Eqs.~(\ref{eq:twistoprel1}) and (\ref{eq:twistoprel2}). We remind the reader that the terms $p^{\{\mu_1} \ldots p^{\mu_n\}}$ and $S^{\{\mu_0} p^{\mu_1} \ldots p^{\mu_m\}}$ arising from the Feynman rules and spinor algebra for the chiral extrapolation of the matrix elements factor out when writing out the quark moment chiral extrapolation (again see Eqs.~(\ref{eq:twistoprel1}) and (\ref{eq:twistoprel2})).

\begin{widetext}

The general expression for the $n \ge 2$ spin-independent moments is

\begin{align}
\nonumber
2\langle x^{n-1} \rangle^B_q = & (C_{BBO_q}^{(n)}+C_{BBO_qM}^{(n)}) + C^{(n)}_{BB'\phi} C^{(n)}_{B'B''O_q} C^{(n)}_{B''B\phi}Z_{1,U}^{(8,8)}(m_\phi^2) +  C^{(n)}_{BB\phi\phi O_q}Z_{1,U}^{\textrm{tad}}(m_\phi^2) \\ \nonumber
 & {} +  C^{(n)}_{BT\phi} C^{(n)}_{TT'O_q} C^{(n)}_{T'B\phi}\left[Z_{1,U1}^{(10,10)}(m_\phi^2)+Z_{1,U2}^{(10,10)}(m_\phi^2)\right]\\& -  (C^{(n)}_{BB'\phi})^2C_{BBO_q}Z_{2,U}^{8}(m_\phi^2) - (C^{(n)}_{BT\phi})^2C^{(n)}_{BBO_q}Z_{2,U}^{10}(m_\phi^2),\\[12pt] \nonumber
\intertext{while the $n=1$ case is simply the quark flavor sum rule. The spin dependent moments are given, for $m\ge0$, by}
\nonumber
2\langle x^m \rangle^B_{\Delta q} = & (C^{(m)}_{BBO_q}+C^{(m)}_{BBO_{\Delta q}M}) + C^{(m)}_{BB'\phi}C^{(m)}_{B'B''O_{\Delta q}}C^{(m)}_{B''B\phi}Z_{1,P}^{(8,8)}(m_\phi^2) +  C^{(m)}_{BB\phi\phi O_{\Delta q}}Z_{1,P}^{\textrm{tad}}(m_\phi^2) \\ \nonumber
 & {} +  C^{(m)}_{BT\phi}C^{(m)}_{TT'O_{\Delta q}}C^{(m)}_{T'B\phi}\left[Z_{1,P1}^{(10,10)}(m_\phi^2) +Z_{1,P2}^{(10,10)}(m_\phi^2) \right] \\ \nonumber
 & {}  +  C^{(m)}_{BT\phi}C^{(m)}_{TB'O_{\Delta q}}C^{(m)}_{B'B\phi}\left[Z_{1,P}^{(8,10)}(m_\phi^2)+Z_{1,P}^{(10,8)}(m_\phi^2)\right] \\ 
 & {} -  (C^{(m)}_{BB'\phi})^2C^{(m)}_{BBO_{\Delta q}}Z_{2,P}^{8}(m_\phi^2) - (C^{(m)}_{BT\phi})^2C_{BBO_{\Delta q}}Z_{2,P}^{10}(m_\phi^2).
\end{align}
%
%
%
%
The term $Z_{1,P2}^{(10,10)}(m_\phi^2)$ contributes only for $m\ge 2$, by Eq.~(\ref{eq:sddec}).
The expressions above match those of previous works~\cite{Arndt:2001ye,Chen:2001gra,Chen:2001eg,Arndt:2001ye} in the limit $\epsilon \rightarrow 0$.

\end{widetext}

\subsubsection{$g_A$ and $\langle x \rangle_{u-d}^p$}

To facilitate direct comparison with and use of these expressions, the chiral expansions for $\langle 1 \rangle^p_{\Delta u - \Delta d} =g_A$ and $\langle x \rangle_{u-d}^p$ are given explicitly.
Again, these expressions match earlier work~\cite{Luty:1993gi,Detmold:2002nf} in the limit $\epsilon \rightarrow 0$. As outlined in previous sections, the integrals $J$ correspond directly in DR to logarithmic contributions of the form $m^2 \textrm{log}(m^2)$. Here the linear terms have been left in terms of the quark masses $Bm_q$. In matching with familiar notation, we identify $\Delta \gamma^{(0)}=2\mathcal{H}$. For our numerical results we impose the SU(6) relation $\mathcal{H}=-3D$.

\begin{widetext}
\begin{equation}
g_A = a + b_{M} + \frac{1}{16 \pi^2 f^2}(d + d' \mathcal{C}^2)
\end{equation}
 %
%
\begin{align}
a = & D + F, \\[10pt] \nonumber
b_M = & \frac{1}{2}\left[\left(-\Delta b^{(0)}_1+\Delta b^{(0)}_2-\Delta b^{(0)}_3+\Delta b^{(0)}_4+\Delta b^{(0)}_5+\Delta b^{(0)}_7\right) Bm_u+\left(-\Delta b^{(0)}_5+\Delta b^{(0)}_7\right)Bm_d \right.\\
& ~~~~+\left.\left(\Delta b^{(0)}_1 + \Delta b^{(0)}_2 + \Delta b^{(0)}_3 + \Delta b^{(0)}_4 + \Delta b^{(0)}_7\right) Bm_s\right], \\[10pt] \nonumber
d = & -\frac{1}{9}(D+F)\left[-3 (D + F)\textrm{cos}\epsilon+ \sqrt{3} (D-3F)\textrm{sin}\epsilon\right]^2\widetilde{J}(m_{\pi_0}^2) \\ \nonumber
& {} -(D+F)\left[(D+F)^2\widetilde{J}(m_{\pi^\pm}^2)+\widetilde{J}_T(m_{\pi^\pm}^2)\right] \\ \nonumber
& {} -\frac{1}{2}(D-F)\left\{\left[2F +3(D+F)\right](D-F)\widetilde{J}(m_{K^0}^2) + \widetilde{J}_T(m_{K^0}^2)\right\}\\ \nonumber
& {}- \frac{1}{3}\left[2D^3+D^2F +12 D F^2 + 9 F^3 \right]\widetilde{J}(m_{K^\pm}^2)-F \widetilde{J}_T(m_{K^\pm}^2) \\
& {} -\frac{1}{9}(D+F)\left[3(D+F)\sin{\epsilon}+\sqrt{3}(D-3F)\textrm{cos}\epsilon\right]^2\widetilde{J}(m_\eta^2), \\[10pt] \nonumber
d' = & -\frac{10}{81}(-3D)\left[(\textrm{cos}^2\epsilon) \widetilde{J}_2(m_{\pi^0}^2,\delta) + 4\widetilde{J}_2(m_{\pi^\pm}^2,\delta)+ \widetilde{J}_2(m_{K^0}^2,\delta)+(\textrm{sin}^2\epsilon) \widetilde{J}_2(m_{\eta}^2,\delta)\right] \\ \nonumber
 & - \frac{1}{6}(D+F)\left[4(\textrm{cos}^2\epsilon) \widetilde{J}_2(m_{\pi^0}^2,\delta) +8\widetilde{J}_2(m_{\pi^\pm}^2,\delta) +2 \widetilde{J}_2(m_{K^0}^2,\delta)+\widetilde{J}_2(m_{K^\pm}^2,\delta)+4(\textrm{sin}^2\epsilon) \widetilde{J}_2(m_{\eta}^2,\delta) \right] \\ \nonumber
& {}+ \frac{2}{9}\left\{4(\textrm{cos}\epsilon) \left[(D+F)\textrm{cos}\epsilon-\frac{1}{\sqrt{3}}(D-3F)\textrm{sin}\epsilon\right]\widetilde{J}_1(m_{\pi^0}^2, \delta) + 4(D+F)\widetilde{J}_1(m_{\pi^\pm}^2, \delta) \right.\\ \nonumber
& {}~~~~~~~ + 2(D-F)\widetilde{J}_1(m_{K^0}^2,\delta)+ (D+3F)\widetilde{J}_1(m_{K\pm}^2,\delta) \\
& {}~~~~~~~ +\left. 4(\textrm{sin}\epsilon) \left[(D+F)\textrm{sin}\epsilon+\frac{1}{\sqrt{3}}(D-3F)\textrm{cos}\epsilon\right]\widetilde{J}_1(m_{\pi^0}^2, \delta)\right\}.
\end{align}
%

%

\begin{equation}
\langle x \rangle_{u-d}^p = \overline{a} + \overline{b}_{M} + \frac{1}{16 \pi^2 f^2}(\overline{d} + \overline{d}' \mathcal{C}^2)
\end{equation}
 %
%
\begin{align}
\overline{a} = & \frac{1}{3}\left( \alpha^{(2)}-\frac{1}{2}\beta^{(2)} \right), \\[10pt] \nonumber
\overline{b}_M = & \frac{1}{2}\left[\left(-b^{(2)}_1+b^{(2)}_2-b^{(2)}_3+b^{(2)}_4+ b^{(2)}_5+ b^{(2)}_7\right) Bm_u+\left(- b^{(2)}_5+ b^{(2)}_7\right)Bm_d \right.\\
& ~~~~+\left.\left( b^{(2)}_1 + b^{(2)}_2 + b^{(2)}_3 + b^{(2)}_4 +  b^{(2)}_7\right) Bm_s\right], \\[10pt] \nonumber
%
\overline{d} = &  -\frac{1}{6}\left(2\alpha^{(2)}-\beta^{(2)}\right)\left[3(D+F)^2\widetilde{J}(m_{\pi^\pm}^2)+\widetilde{J}_T(m_{\pi^\pm}^2)\right] \\ \nonumber
& {} +\frac{1}{24}\left(\alpha^{(2)}+4 \beta^{(2)}\right) \left[ 3(D-F)^2 \widetilde{J}(m_{K^0}^2) + 2 \widetilde{J}_T(m_{K^0}^2)\right]\\ \nonumber
& {}- \frac{1}{24}\left[\left\{6DF\left(\alpha^{(2)}-2\beta^{(2)}\right)+3F^2\left(\alpha^{(2)}+2\beta^{(2)}\right)+D^2\left(11\alpha^{(2)}-10\beta^{(2)}\right) \right\}\widetilde{J}(m_{K^\pm}^2) \right.\\
& {}~~~~~~~~ \left. +\left(5\alpha^{(2)}+2\beta^{(2)}\right) \widetilde{J}_T(m_{K^\pm}^2)\right] \\[10pt] \nonumber
\overline{d}' = & -\frac{1}{9}\left( \gamma^{(2)}-\gamma'^{(2)} \right)\left[(\textrm{cos}^2\epsilon) \widetilde{J}_2(m_{\pi^0}^2,\delta) + 4\widetilde{J}_2(m_{\pi^\pm}^2,\delta)+ \widetilde{J}_2(m_{K^0}^2,\delta)+(\textrm{sin}^2\epsilon) \widetilde{J}_2(m_{\eta}^2,\delta)\right] \\ 
 & - \frac{1}{36}\left(2\alpha^{(2)}-\beta^{(2)}\right)\left[4(\textrm{cos}^2\epsilon) \widetilde{J}_2(m_{\pi^0}^2,\delta) +8\widetilde{J}_2(m_{\pi^\pm}^2,\delta) +2 \widetilde{J}_2(m_{K^0}^2,\delta)+\widetilde{J}_2(m_{K^\pm}^2,\delta) +4(\textrm{sin}^2\epsilon) \widetilde{J}_2(m_{\eta}^2,\delta) \right]
\end{align}

\end{widetext}

\begin{figure}[htbf]
\centering
\subfigure[Ratio of singly-represented quark moments.]{
\includegraphics[width=\columnwidth]{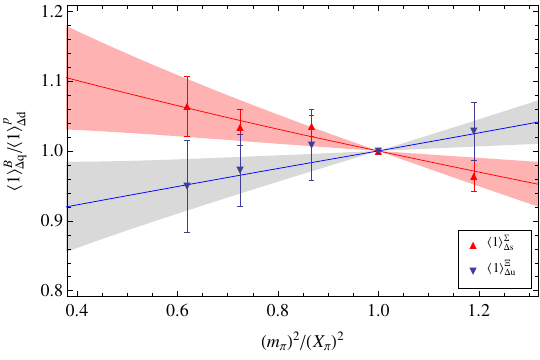}
\label{fig:ZeroSDSingly}
}
\subfigure[Ratio of doubly-represented quark moments.]{
\centering
\includegraphics[width=\columnwidth]{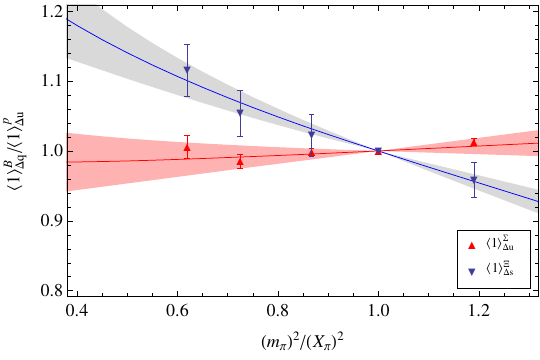}
\label{fig:ZeroSDDoubly}
}
\caption{Illustration of the fit to the zeroth spin-dependent moments -- data from Ref.~\cite{Horsley:2010th,Cloet:2012db}.}
\label{fig:SD0}
\end{figure}

\begin{figure}[htbf]
\centering
\subfigure[Ratio of singly-represented quark moments.]{
\includegraphics[width=\columnwidth]{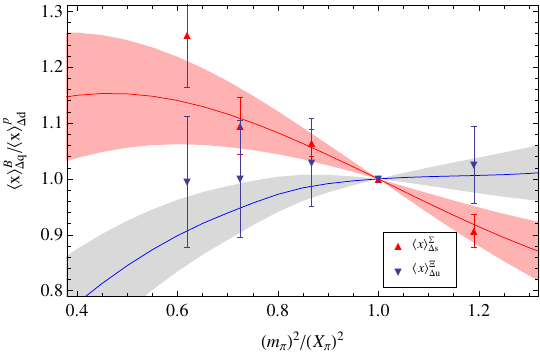}
\label{fig:FirstSDSingly}
}
\subfigure[Ratio of doubly-represented quark moments.]{
\centering
\includegraphics[width=\columnwidth]{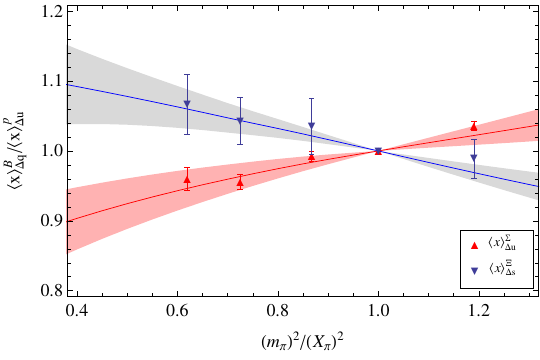}
\label{fig:FirstSDDoubly}
}
\caption{Illustration of the fit to the first spin-dependent moments -- data from Ref.~\cite{Horsley:2010th,Cloet:2012db}.}
\label{fig:SD1}
\end{figure}

\begin{figure}[htbf]
\centering
\subfigure[Ratio of singly-represented quark moments.]{
\includegraphics[width=\columnwidth]{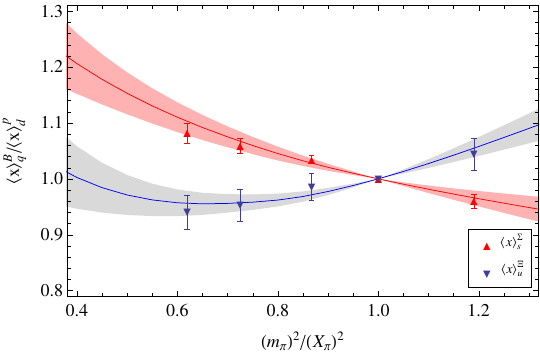}
\label{fig:FristSISingly}
}
\subfigure[Ratio of doubly-represented quark moments.]{
\centering
\includegraphics[width=\columnwidth]{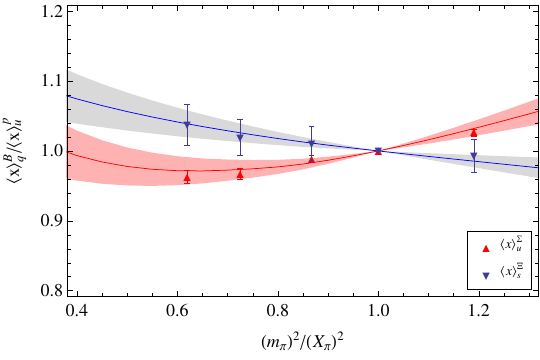}
\label{fig:FirstSIDoubly}
}
\caption{Illustration of the fit to the first spin-independent moments -- data from Ref.~\cite{Horsley:2010th,Cloet:2012db}.}
\label{fig:SI1}
\end{figure}

\begin{figure}[htbf]
\centering
\includegraphics[width=\columnwidth]{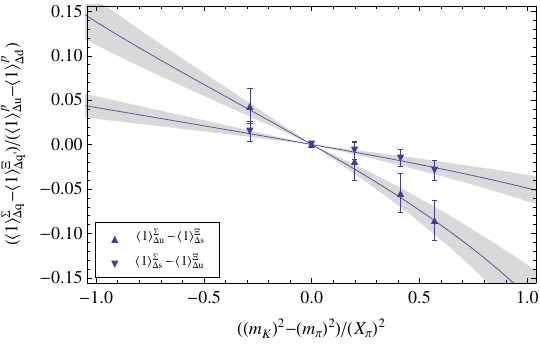}
\caption{Illustration of the fit to the zeroth spin-dependent moments -- data from Ref.~\cite{Horsley:2010th,Cloet:2012db}.}
\label{fig:ZeroSD}
\end{figure}

\begin{figure}[htbf]
\centering
\includegraphics[width=\columnwidth]{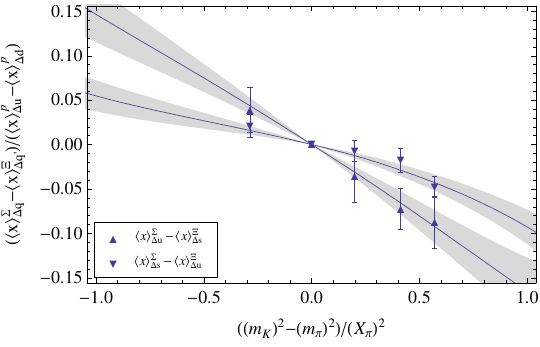}
\caption{Illustration of the fit to the first spin-dependent moments -- data from Ref.~\cite{Horsley:2010th,Cloet:2012db}.}
\label{fig:FirstSD}
\end{figure}

\begin{figure}[htbf]
\centering
\includegraphics[width=\columnwidth]{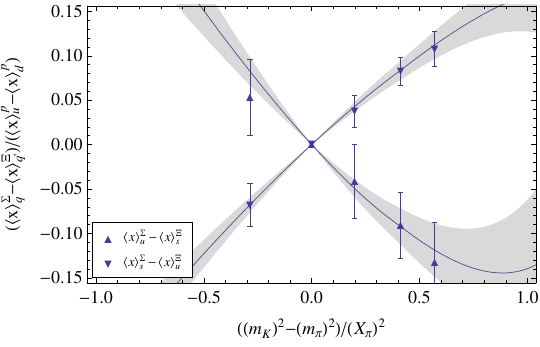}
\caption{Illustration of the fit to the first spin-independent moments -- data from Ref.~\cite{Horsley:2010th,Cloet:2012db}.}
\label{fig:FirstSI}
\end{figure}

\section{Chiral extrapolation of lattice data}
\label{sec:latticefits}

In this section we describe the application of the theory developed here to the chiral extrapolation of lattice results provided by the CSSM and QCDSF/UKQCD Collaborations for the first few Mellin moments of the quark distributions~\cite{Horsley:2010th,Cloet:2012db,private,Bietenholz:2011qq}. In particular, we consider the first spin-independent moment and the zeroth and first spin-dependent moments. We emphasize that the fits shown involve only published results~\cite{Horsley:2010th,Cloet:2012db}, and are intended as merely an illustration of the applicability of this work; ideally a full quantitative analysis should involve additional lattice results and account for correlations between the data points.

We choose to use a dipole regulator $u(k)=\left(\frac{\Lambda^2}{\Lambda^2+k^2}\right)^2$ and a regulator mass $\Lambda = 1$~GeV within the FRR scheme. Our results are insensitive to this choice; choosing different regulator forms, for example monopole, Gaussian or sharp cutoff, and allowing $\Lambda$ to vary by $\pm 20 \%$ does not change the results of the analysis within the quoted uncertainties.

The fit to the lattice results is performed by minimizing the sum of $\chi^2$ for each set of moments. As data is available only for the doubly- and singly-represented quark moments~\cite{private}, not all of the parameters which appear in the previous sections are linearly independent in the relevant fit functions. 
Replacements are made:
\begin{align}
n_1 & = b_1 + b_3 & n_2 & = b_2 + b_4 & n_3 & = b_5 \\
n_4 & = b_7 & n_5 & = b_8 & n_6 & = b_9,
\end{align}
with entirely analogous substitutions giving $\Delta n_i$ in the spin-dependent cases.

The fit parameters are different for each of the three moments under consideration. In each case we use SU(6) relations between unknown quantities to reduce the number of free parameters. There are 24 lattice data points available for each moment considered~\cite{private}. 
\begin{itemize}
\item For the zeroth spin-dependent moment, $\Delta n_i^{(0)}$, $D$, and $\Delta\sigma^{(0)}$ are fit, with SU(6) symmetry used to set $F=\frac{2}{3}D$ and $\Delta \gamma^{(0)} = -6D$. $\mathcal{C}\rightarrow \mathcal{C}_{\textrm{phys}} = -\frac{6}{5}g_{A_{\textrm{phys}}}$ is also fixed.  In this case, there are eight free parameters.
\item The nine fit parameters for the first spin-dependent moment are $\Delta n_i^{(1)}$, $\Delta \alpha^{(1)}$, $\Delta \beta^{(1)}$ and $\Delta \sigma^{(1)}$. Fixed parameters are $D \rightarrow D_{\textrm{phys}}=\frac{3}{5}g_{A_{\textrm{phys}}}$, $F \rightarrow F_{\textrm{phys}}=\frac{2}{3}D_{\textrm{phys}}$, $\mathcal{C}\rightarrow \mathcal{C}_{\textrm{phys}}$, and, using SU(6) symmetry, $\Delta \gamma^{(1)}=-\frac{3}{2}(\Delta \alpha^{(1)}-2\Delta \beta^{(1)})$ as outlined in the text.
\item For the first spin-independent moment, nine parameters, $n_i^{(2)}$, $\alpha^{(2)}$, $\beta^{(2)}$ and $\sigma^{(2)}$, are fit, with $D$, $F$ and $\mathcal{C}$ again fixed to their physical values. As no phenomenological estimate of this quantity is available, the combination $(\gamma^{(2)} - \frac{\gamma'^{(2)}}{3})$ is fixed to a `physical' value; using the experimental tree level delta insertion as input~\cite{Martin:2002aw}  
\begin{align}
(\gamma^{(2)} - \frac{\gamma'^{(2)}}{3}) & = 6\langle x \rangle^{\Delta^+}_{u-d} \hspace{1cm} \textrm{at tree level} \\
 & = 6\langle x \rangle^p_{u-d} \\
 & \underset{\tiny{phys}}{=} 6(0.157) = 0.942.
\end{align} 
\end{itemize}

The fits are shown in Figs.~\ref{fig:SD0}, \ref{fig:SD1}, \ref{fig:SI1}. Here $X_\pi = \sqrt{(2m_K^2+m_\pi^2)/3}=411$~MeV is the simulation centre-of-mass of the pseudoscalar meson octet. Ratios of moments are displayed and the $X_\pi$ normalization is taken for the figures so that they may be easily compared against published results~\cite{Horsley:2010th,*Cloet:2012db}. The quality of fit is clearly acceptable in each case with $\chi^2/\textrm{dof}$ between 0.6 and 0.9 for each moment. All $\chi^2$ values are less than one as we were not able to take into account the effect of correlations between the original lattice data. Best-fit parameters are shown in Table~\ref{tab:app}.

\begin{table*}[tbh]
\begin{center}
\begin{ruledtabular}
\begin{tabular}{lccccccccc}
first SI & $n_1^{(2)}$ & $n_2^{(2)}$  & $n_3^{(2)}$ & $n_4^{(2)}$  & $n_5^{(2)}$  & $n_6^{(2)}$  & $\alpha^{(2)}$ & $\beta^{(2)}$ & $\sigma^{(2)}$  \\ 
 & 1.1(25)(0) & -7.0(28)(27) & 8.3(26)(31) & 0.5(27)(1) & 11(4)(4) & 6.2(24)(23) & -4.1(17)(12) & -8.6(31)(21) & 7.5(26)(23)  \\ \hline

zeroth SD & $\Delta n_1^{(0)}$  & $\Delta n_2^{(0)}$  & $\Delta n_3^{(0)}$  & $\Delta n_4^{(0)}$  & $\Delta n_5^{(0)}$  & $\Delta n_6^{(0)}$ & $D$ & $\Delta \sigma^{(0)}$  &\\ 
 & 4.9(84)(9) & 0.5(98)(12) & -2.2(58)(9) & -15(17)(0) & 0.2(50)(9) & -1.1(88)(7) & 0.74(24)(6) & -0.22(26)(0) &   \\
\hline
first SD & $\Delta n_1^{(1)}$  & $\Delta n_2^{(1)}$  & $\Delta n_3^{(1)}$  & $\Delta n_4^{(1)}$  & $\Delta n_5^{(1)}$  & $\Delta n_6^{(1)}$  & $\Delta \alpha^{(1)}$ & $\Delta \beta^{(1)}$ & $\Delta \sigma^{(1)}$  \\ 
 & -1.5(13)(15) & 6.3(29)(26) & -3.9(16)(23) & -7.0(46)(11) & -1.0(11)(8) & -6.0(28)(34) & 0.41(50)(29) & -1.5(10)(3) & -0.93(61)(14) \\
\end{tabular}

\end{ruledtabular}
\end{center}
\caption{Values of the fit parameters corresponding to the fits shown in Figs.~\ref{fig:SD0},~\ref{fig:SD1} and \ref{fig:SI1}. All $(\Delta)n^{(j)}_i$ have dimensions (GeV$^{-2}$), other parameters are dimensionless. The first uncertainty given is statistical, while the second indicates the uncertainty resulting from a $\pm$20\% variation in the FRR cutoff $\Lambda$. }
\label{tab:app}
\end{table*}

\section{Conclusion}

We have developed chiral extrapolation formulae for the matrix elements of local twist-2 operators including the effects of isospin breaking. From these, we infer similar formulae for the chiral extrapolation of spin-dependent and spin-independent moments of quark distribution functions. The analysis includes loop corrections and counterterms to leading non-analytic order.
This work represents an extension of previous results in that we allow for a non-zero light-quark mass difference. This allows our results to be used for the chiral extrapolation of both (2+1) and (1+1+1)-flavor lattice results to the physical point. Such lattice results may then be directly compared with experimental values. In section~\ref{sec:latticefits} we presented an example of such an application to the results of recent lattice simulations.

We emphasize that the application presented here is merely an illustration, with the fits performed to a limited amount of data. The true usefulness of our analysis and technique will come from the facility to extrapolate to the physical point. When the results of more lattice simulations become publicly available, in particular for quark distribution moments with lattice-determined normalizations, rather than in ratio form, the extrapolations developed here will allow a valuable comparison of lattice data with experimental results at the physical point.

\section*{Acknowledgements}
We gratefully acknowledge the assistance of the QCDSF/UKQCD Collaboration in providing access to results which are not yet published. We also acknowledge helpful discussions with J.~Zanotti and W.~Detmold.
This work was supported by the University of Adelaide and the Australian
Research Council through through the ARC Centre of Excellence for Particle Physics at the Terascale and grants FL0992247 (AWT) and DP110101265 (RDY) and FT120100821 (RDY).

\FloatBarrier
\clearpage


\appendix



\begin{widetext}




\section{Coefficient tables}
\label{app:coeffs}

Superscripts $^{(n)}$ may be assumed on every Clebsch-Gordan coefficient $C$ and on every unknown parameter $\alpha$ etc. These tables are identical for the spin-dependent case, for which all unknown parameters may be substituted, for example $\alpha^{(n)} \rightarrow \Delta \alpha^{(n)}$.

\begin{table}[htbf]
\begin{centering}
\begin{ruledtabular}
\begin{tabular}{ccccccccc}
\multicolumn{9}{c}{$C_{BB'O_u}$}\\
\backslashbox{$B'$}{$B$}&$p$ & $n$ & $\Lambda$ & $\Sigma^0$ & $\Sigma^+$ & $\Sigma^-$ & $\Xi^0$ & $\Xi^-$ \\
\hline
$p$& $\frac{5 \alpha  }{6}+\frac{\beta  }{3}+\sigma  $ &   &   &   &   &   &   &   \\
$n$&   & $\frac{1}{6} \left(\alpha  +4 \beta  +6 \sigma  \right)$ &   &   &   &   &   &   \\
$\Lambda$ &  &   & $\frac{1}{4} \left(\alpha  +2 \beta  +4 \sigma  \right)$ & $\frac{\alpha  -2 \beta  }{4 \sqrt{3}}$ &   &   &   &   \\
$\Sigma^0$&   &   & $\frac{\alpha  -2 \beta  }{4 \sqrt{3}}$ & $\frac{5 \alpha  }{12}+\frac{\beta  }{6}+\sigma  $ &   &   &   &   \\
$\Sigma^+$ &  &   &   &   & $\frac{5 \alpha  }{6}+\frac{\beta  }{3}+\sigma  $ &   &   &   \\
$\Sigma^-$&   &   &   &   &   & $\sigma  $ &   &   \\
$\Xi^0$&   &   &   &   &   &   & $\frac{1}{6} \left(\alpha  +4 \beta  +6 \sigma  \right)$ &   \\
$\Xi^-$&   &   &   &   &   &   &   & $\sigma  $ 
\end{tabular}
\end{ruledtabular}
\end{centering}
\end{table}

\begin{table}[htbf]
\begin{centering}
\begin{ruledtabular}
\begin{tabular}{ccccccccc}
\multicolumn{9}{c}{$C_{BB'O_d}$}\\
\backslashbox{$B'$}{$B$}&$p$ & $n$ & $\Lambda$ & $\Sigma^0$ & $\Sigma^+$ & $\Sigma^-$ & $\Xi^0$ & $\Xi^-$ \\
\hline
$p$& $\frac{1}{6} (\alpha +4 \beta +6 \sigma )$ &   &   &   &   &   &   &   \\
$n$&   & $\frac{5 \alpha }{6}+\frac{\beta }{3}+\sigma$  &   &   &   &   &   &   \\
$\Lambda$&   &   & $\frac{1}{4} (\alpha +2 \beta +4 \sigma )$ & $-\frac{\alpha -2 \beta }{4 \sqrt{3}}$ &   &   &   &   \\
$\Sigma^0$&    &   & $-\frac{\alpha -2 \beta }{4 \sqrt{3}}$ & $\frac{5 \alpha }{12}+\frac{\beta }{6}+\sigma$  &   &   &   &   \\
$\Sigma^+$ &   &   &   &   & $\sigma$  &   &   &   \\
$\Sigma^-$&    &   &   &   &   & $\frac{5 \alpha }{6}+\frac{\beta }{3}+\sigma$  &   &   \\
$\Xi^0$&   &   &   &   &   &   & $\sigma$  &   \\
$\Xi^-$&   &   &   &   &   &   &  & $\frac{1}{6} (\alpha +4 \beta +6 \sigma )$ \\
\end{tabular}
\end{ruledtabular}
\end{centering}
\end{table}

\begin{table}[htbf]
\begin{centering}
\begin{ruledtabular}
\begin{tabular}{ccccccccc}
\multicolumn{9}{c}{$C_{BB'O_s}$}\\
\backslashbox{$B'$}{$B$}&$p$ & $n$ & $\Lambda$ & $\Sigma^0$ & $\Sigma^+$ & $\Sigma^-$ & $\Xi^0$ & $\Xi^-$ \\
\hline
$p$& $\sigma$  &   &   &   &   &   &   &   \\
$n$&   & $\sigma$  &   &   &   &   &   &   \\
$\Lambda$&   &   & $\frac{\alpha }{2}+\sigma$  &   &   &   &   &   \\
$\Sigma^0$&   &   &   & $\frac{1}{6} (\alpha +4 \beta +6 \sigma )$ &   &   &   &   \\
$\Sigma^+$ &   &   &   &   & $\frac{1}{6} (\alpha +4 \beta +6 \sigma )$ &   &   &   \\
$\Sigma^-$&   &   &   &   &   & $\frac{1}{6} (\alpha +4 \beta +6 \sigma )$ &   &   \\
$\Xi^0$&   &   &   &   &   &   & $\frac{5 \alpha }{6}+\frac{\beta }{3}+\sigma$  &   \\
$\Xi^-$&   &   &   &   &   &   &   & $\frac{5 \alpha }{6}+\frac{\beta }{3}+\sigma$  \\
\end{tabular}
\end{ruledtabular}
\end{centering}
\end{table}

\clearpage

\begin{table}
\begin{centering}
\begin{ruledtabular}
\begin{tabular}{cccc}
\multicolumn{4}{c}{$C_{BBO_uM}$} \\
\backslashbox{$B$}{} & $\times m_u^{-1}$ & $\times m_d^{-1}$ & $\times m_s^{-1}$\\ \hline
$p$& -$b_1+b_2-b_3+b_4+b_5+b_7+b_9$ & $b_7$ & $b_1+b_2+b_3+b_4+b_7+b_8$ \\
$n$& $b_5$ & $b_9$ & $b_8$ \\
$\Lambda$&  $\frac{1}{6} (b_{10}+4b_4+6b_5+b_6+b_7+b_8+b_9)$ & $\frac{1}{6} (b_{10}+b_6+b_7+b_8+b_9)$ & $\frac{1}{6} (-2 b_{10}+b_6+b_7+4b_8+4b_9)$ \\
$\Sigma^0$&  $\frac{1}{2} (b_{10}+4 b_4+2 b_5+b_6+b_7+b_8+b_9)$ & $\frac{1}{2} (-b_{10}+b_6+b_7+b_8+b_9)$ & $\frac{b_6+b_7}{2}$ \\
$\Sigma^+$ & $-b_1+b_2-b_3+b_4+b_5+b_7+b_9$ & $b_1+b_2+b_3+b_4+b_7+b_8$ & $b_7$ \\
$\Sigma^-$& $-b_1-b_2+b_3+b_4+b_5+b_6+b_8$ & $b_1-b_2-b_3+b_4+b_6+b_9$ & $b_6$ \\
$\Xi^0$& $b_5$ & $b_8$ & $b_9$ \\
$\Xi^-$& $-b_1-b_2+b_3+b_4+b_5+b_6+b_8$ & $b_6$ & $b_1-b_2-b_3+b_4+b_6+b_9$ \\
\end{tabular}
\end{ruledtabular}
\end{centering}
\end{table}

\begin{table}[htbf]
\begin{centering}
\begin{ruledtabular}
\begin{tabular}{cccc}
\multicolumn{4}{c}{$C_{BBO_dM}$} \\
\backslashbox{$B$}{} & $\times m_u^{-1}$ & $\times m_d^{-1}$ & $\times m_s^{-1}$\\ \hline
$p$& $b_9$ & $b_5$ & $b_8$ \\
$n$& $b_7$ & $-b_1+b_2-b_3+b_4+b_5+b_7+b_9$ & $b_1+b_2+b_3+b_4+b_7+b_8$ \\
$\Lambda$& $\frac{1}{6} (b_{10}+b_6+b_7+b_8+b_9)$ & $\frac{1}{6} (b_{10}+4 b_4+6 b_5+b_6+b_7+b_8+b_9)$ & $\frac{1}{6} (-2b_{10}+b_6+b_7+4 b_8+4 b_9)$ \\
$\Sigma^0$ & $\frac{1}{2} (-b_{10}+b_6+b_7+b_8+b_9)$ & $\frac{1}{2} (b_{10}+4 b_4+2 b_5+b_6+b_7+b_8+b_9)$ & $\frac{b_6+b_7}{2}$ \\
$\Sigma^+$ & $b_1-b_2-b_3+b_4+b_6+b_9$ & $-b_1-b_2+b_3+b_4+b_5+b_6+b_8$ & $b_6$ \\
$\Sigma^-$& $b_1+b_2+b_3+b_4+b_7+b_8$ & $-b_1+b_2-b_3+b_4+b_5+b_7+b_9$ & $b_7$ \\
$\Xi^0$& $b_6$ & $-b_1-b_2+b_3+b_4+b_5+b_6+b_8$ & $b_1-b_2-b_3+b_4+b_6+b_9$ \\
$\Xi^-$& $b_8$ & $b_5$ & $b_9$ \\
\end{tabular}
\end{ruledtabular}
\end{centering}
\end{table}

\begin{table}[htbf]
\begin{centering}
\begin{ruledtabular}
\begin{tabular}{cccc}
\multicolumn{4}{c}{$C_{BBO_sM}$} \\
\backslashbox{$B$}{} & $\times m_u^{-1}$ & $\times m_d^{-1}$ & $\times m_s^{-1}$\\ \hline
$p$& $b_1-b_2-b_3+b_4+b_6+b_9$ & $b_6$ & $-b_1-b_2+b_3+b_4+b_5+b_6+b_8$ \\
$n$& $b_6$ & $b_1-b_2-b_3+b_4+b_6+b_9$ & $-b_1-b_2+b_3+b_4+b_5+b_6+b_8$ \\
$\Lambda$& $\frac{1}{6} (-2 b_{10}+4 b_6+4 b_7+b_8+b_9)$ & $\frac{1}{6} (-2 b_{10}+4 b_6+4 b_7+b_8+b_9)$ & $\frac{1}{3} (2 b_{10}+8 b_4+3 b_5+2b_6+2 b_7+2 b_8+2 b_9)$ \\
$\Sigma^0$& $\frac{b_8+b_9}{2}$ & $\frac{b_8+b_9}{2}$ & $b_5$ \\
$\Sigma^+$ & $b_9$ & $b_8$ & $b_5$ \\
$\Sigma^-$& $b_8$ & $b_9$ & $b_5$ \\
$\Xi^0$& $b_7$ & $b_1+b_2+b_3+b_4+b_7+b_8$ & $-b_1+b_2-b_3+b_4+b_5+b_7+b_9$ \\
$\Xi^-$& $b_1+b_2+b_3+b_4+b_7+b_8$ & $b_7$ & $-b_1+b_2-b_3+b_4+b_5+b_7+b_9$ \\
\end{tabular}
\end{ruledtabular}
\end{centering}
\end{table}

\clearpage

\begin{table}[htbf]
\begin{centering}
\begin{ruledtabular}
\begin{tabular}{ccccc}
\multicolumn{5}{c}{$C_{BB'\pi^0}$}\\
\backslashbox{$B'$}{$B$}&\multicolumn{2}{c}{$p$} & \multicolumn{2}{c}{$n$}   \\
\hline
$p$& \multicolumn{2}{c}{$\frac{1}{3} \sqrt{2} \left(3 (D+F) \cos \epsilon-\sqrt{3} (D-3 F) \sin \epsilon\right)$}  &  &    \\
$n$&  & & \multicolumn{2}{c}{$-\frac{1}{3} \sqrt{2} \left(3 (D+F) \cos \epsilon+\sqrt{3} (D-3 F) \sin \epsilon\right)$}    \\
$\Lambda$&  &  & &   \\
$\Sigma^0$&  &  & & \\
$\Sigma^+$&  &  &  &    \\
$\Sigma^-$&  &  &  &    \\
$\Xi^0 $&  &  &  &   \\
$\Xi^-$&  &  &  &    \\
\hline
\backslashbox{$B'$}{$B$}& $\Lambda$ & $\Sigma^0$ & $\Sigma^+$ & $\Sigma^-$ \\
\hline
$p$&  &  &  &   \\
$n$&  &  &  &  \\
$\Lambda$&  $-2 \sqrt{\frac{2}{3}} D \sin \epsilon$ & $2 \sqrt{\frac{2}{3}} D \cos \epsilon$  &  &  \\
$\Sigma^0$&  $2 \sqrt{\frac{2}{3}} D \cos \epsilon$ & $2 \sqrt{\frac{2}{3}} D \sin \epsilon$   &  &  \\
$\Sigma^+$& & & $2 \sqrt{2} F \cos \epsilon+2 \sqrt{\frac{2}{3}} D \sin \epsilon$ &  \\
$\Sigma^-$&  & & &$2 \sqrt{\frac{2}{3}} D \sin \epsilon-2 \sqrt{2} F \cos \epsilon$   \\
$\Xi^0 $&  &  &&  \\
$\Xi^-$&  &  &  &  \\
\hline
\backslashbox{$B'$}{$B$} & \multicolumn{2}{c}{$\Xi^0$} & \multicolumn{2}{c}{$\Xi^-$}\\
\hline
$p$& & & &   \\
$n$&  &  &  &  \\
$\Lambda$&  &  &  &  \\
$\Sigma^0$&  &  &  &  \\
$\Sigma^+$& &  &  &  \\
$\Sigma^-$&  & &  &  \\
$\Xi^0 $&  \multicolumn{2}{c}{$-\frac{1}{3} \sqrt{2} \left(3 (D-F) \cos \epsilon+\sqrt{3} (D+3 F) \sin \epsilon\right)$} & & \\
$\Xi^-$&  &  &  \multicolumn{2}{c}{$\frac{1}{3} \sqrt{2} \left(3 (D-F) \cos \epsilon-\sqrt{3} (D+3 F) \sin \epsilon\right)$} \\
\end{tabular}
\end{ruledtabular}
\end{centering}
\end{table}



\begin{table}
\begin{centering}
\begin{ruledtabular}
\begin{tabular}{ccccccccc}
\multicolumn{9}{c}{$C_{BB\pi^+}$}\\
&$p$ & $n$ & $\Lambda$ & $\Sigma^0$ & $\Sigma^+$ & $\Sigma^-$ & $\Xi^0$ & $\Xi^-$ \\
\hline
 $p$&   &   &   &   &   &   &   &   \\
  $n$& $2 (D+F)$ &   &   &   &   &   &   &   \\
 $\Lambda$&    &   &   &   & $2 \sqrt{\frac{2}{3}}D$ &   &   &   \\
  $\Sigma^0$&  &   &   &   & $-2 \sqrt{2} F$ &   &   &   \\
  $\Sigma^+$ &  &   &   &   &   &   &   &   \\
 $\Sigma^-$&   &   & $2 \sqrt{\frac{2}{3}} D$ & $2 \sqrt{2} F$ &   &   &   &   \\
 $\Xi^0 $&  &   &   &   &   &   &   &   \\
 $\Xi^-$&  &   &   &   &   &   & $2( D-F)$ &   \\
\end{tabular}
\end{ruledtabular}
\end{centering}
\end{table}

\begin{table}[htbf]
\begin{centering}
\begin{ruledtabular}
\begin{tabular}{ccccccccc}
\multicolumn{9}{c}{$C_{BB'\pi^-}$}\\
\backslashbox{$B'$}{$B$}&$p$ & $n$ & $\Lambda$ & $\Sigma^0$ & $\Sigma^+$ & $\Sigma^-$ & $\Xi^0$ & $\Xi^-$ \\
\hline
 $p$&  & $2 (D+F)$ &  &  &  &  &  &  \\
$n$&  &  &  &  &  &  &  &  \\
 $\Lambda$&  &  &  &  &  & $2 \sqrt{\frac{2}{3}} D$ &  &  \\
   $\Sigma^0$& &  &  &  &  & $2 \sqrt{2} F$ &  &  \\
  $\Sigma^+$ &  &  & $2 \sqrt{\frac{2}{3}} D$ & $-2 \sqrt{2} F$ &  &  &  &  \\
 $\Sigma^-$&  &  &  &  &  &  &  &  \\
  $\Xi^0 $& &  &  &  &  &  &  & $2 (D-F)$ \\
 $\Xi^-$&  &  &  &  &  &  &  &  \\
\end{tabular}
\end{ruledtabular}
\end{centering}
\end{table}

\begin{table}[htbf]
\begin{centering}
\begin{ruledtabular}
\begin{tabular}{ccccccccc}
\multicolumn{9}{c}{$C_{BB'K^0}$}\\
\backslashbox{$B'$}{$B$}&$p$ & $n$ & $\Lambda$ & $\Sigma^0$ & $\Sigma^+$ & $\Sigma^-$ & $\Xi^0$ & $\Xi^-$ \\
\hline
$p$&  &  &  &  &  &  &  &  \\
$n$&  &  &  &  &  &  &  &  \\
$\Lambda$&  & $-\sqrt{\frac{2}{3}} (D+3F)$ &  &  &  &  &  &  \\
$\Sigma^0$&  & $\sqrt{2} (F-D)$ &  &  &  &  &  &  \\
$\Sigma^+$&$2 (D-F)$ &  &  &  &  &  &  &  \\
$\Sigma^-$&  &  &  &  &  &  &  &  \\
$\Xi^0 $&  &  & $-\sqrt{\frac{2}{3}} (D-3F)$ & $-\sqrt{2} (D+F)$ &  &  &  &  \\
$\Xi^-$&  &  &  &  &  & $2 (D+F)$ &  &  \\
\end{tabular}
\end{ruledtabular}
\end{centering}
\end{table}

\begin{table}[htbf]
\begin{centering}
\begin{ruledtabular}
\begin{tabular}{ccccccccc}
\multicolumn{9}{c}{$C_{BB'K^+}$}\\
\backslashbox{$B'$}{$B$}&$p$ & $n$ & $\Lambda$ & $\Sigma^0$ & $\Sigma^+$ & $\Sigma^-$ & $\Xi^0$ & $\Xi^-$ \\
\hline
$p$&  &  &  &  &  &  &  &  \\
$n$&  &  &  &  &  &  &  &  \\
$\Lambda$& $-\sqrt{\frac{2}{3}} (D+3F)$ &  &  &  &  &  &  &  \\
$\Sigma^0$& $\sqrt{2} (D-F)$ &  &  &  &  &  &  &  \\
$\Sigma^+$&  &  &  &  &  &  &  &  \\
$\Sigma^-$&  & $2 (D-F)$ &  &  &  &  &  &  \\
$\Xi^0 $&  &  &  &  & $2 (D+F)$ &  &  &  \\
$\Xi^-$&  &  & $-\sqrt{\frac{2}{3}} (D-3F)$ & $\sqrt{2} (D+F)$ &  &  &  &  \\
\end{tabular}
\end{ruledtabular}
\end{centering}
\end{table}

\begin{table}[htbf]
\begin{centering}
\begin{ruledtabular}
\begin{tabular}{ccccccccc}
\multicolumn{9}{c}{$C_{BB'K^-}$}\\
\backslashbox{$B'$}{$B$}&$p$ & $n$ & $\Lambda$ & $\Sigma^0$ & $\Sigma^+$ & $\Sigma^-$ & $\Xi^0$ & $\Xi^-$ \\
\hline
$p$&  &  & $-\sqrt{\frac{2}{3}} (D+3F)$ & $\sqrt{2} (D-F)$ &  &  &  &  \\
$n$&  &  &  &  &  & $2 (D-F)$ &  &  \\
$\Lambda$&  &  &  &  &  &  &  & $-\sqrt{\frac{2}{3}} (D-3F)$ \\
$\Sigma^0$&  &  &  &  &  &  &  & $\sqrt{2} (D+F)$ \\
$\Sigma^+$&  &  &  &  &  &  & $2 (D+F)$ &  \\
$\Sigma^-$&  &  &  &  &  &  &  &  \\
$\Xi^0 $&  &  &  &  &  &  &  &  \\
$\Xi^-$&  &  &  &  &  &  &  &  \\
\end{tabular}
\end{ruledtabular}
\end{centering}
\end{table}

\begin{table}[htbf]
\begin{centering}
\begin{ruledtabular}
\begin{tabular}{ccccccccc}
\multicolumn{9}{c}{$C_{BB'\overline{K}^0}$}\\
\backslashbox{$B'$}{$B$}&$p$ & $n$ & $\Lambda$ & $\Sigma^0$ & $\Sigma^+$ & $\Sigma^-$ & $\Xi^0$ & $\Xi^-$ \\
\hline
$p$&  &  &  &  & $2 (D-F)$ &  &  &  \\
$n$&  &  & $-\sqrt{\frac{2}{3}} (D+3F)$ & $\sqrt{2} (F-D)$ &  &  &  &  \\
$\Lambda$&  &  &  &  &  &  & $-\sqrt{\frac{2}{3}} D-3F)$ &  \\
$\Sigma^0$&  &  &  &  &  &  & $-\sqrt{2} (D+F)$ &  \\
$\Sigma^+$&  &  &  &  &  &  &  &  \\
$\Sigma^-$&  &  &  &  &  &  &  & $2 (D+F)$ \\
$\Xi^0 $&  &  &  &  &  &  &  &  \\
$\Xi^-$&  &  &  &  &  &  &  &  \\
\end{tabular}
\end{ruledtabular}
\end{centering}
\end{table}

\begin{table}[htbf]
\begin{centering}
\begin{ruledtabular}
\begin{tabular}{ccccc}
\multicolumn{5}{c}{$C_{BB'\eta}$}\\
\backslashbox{$B'$}{$B$}&\multicolumn{2}{c}{$p$} & \multicolumn{2}{c}{$n$}   \\
\hline
$p$& \multicolumn{2}{c}{$-\frac{1}{3} \sqrt{2} \left(\sqrt{3} (D-3 F) \cos \epsilon+3 (D+F) \sin \epsilon\right)$}  &  &    \\
$n$&  & & \multicolumn{2}{c}{$\sqrt{2} (D+F) \sin \epsilon-\sqrt{\frac{2}{3}} (D-3 F) \cos \epsilon$}    \\
$\Lambda$&  &  & &   \\
$\Sigma^0$&  &  & & \\
$\Sigma^+$&  &  &  &    \\
$\Sigma^-$&  &  &  &    \\
$\Xi^0 $&  &  &  &   \\
$\Xi^-$&  &  &  &    \\
\hline
\backslashbox{$B'$}{$B$}& $\Lambda$ & $\Sigma^0$ & $\Sigma^+$ & $\Sigma^-$ \\
\hline
$p$&  &  &  &   \\
$n$&  &  &  &  \\
$\Lambda$& $-2 \sqrt{\frac{2}{3}} D \cos \epsilon$ & $-2 \sqrt{\frac{2}{3}} D \sin \epsilon$  &  &  \\
$\Sigma^0$&  $-2 \sqrt{\frac{2}{3}} D \sin \epsilon$ & $2 \sqrt{\frac{2}{3}} D \cos \epsilon$  &  &  \\
$\Sigma^+$& & & $2 \sqrt{\frac{2}{3}} D \cos \epsilon-2 \sqrt{2} F \sin \epsilon$ &  \\
$\Sigma^-$&  & & & $2 \sqrt{\frac{2}{3}} D \cos \epsilon+2 \sqrt{2} F \sin \epsilon$   \\
$\Xi^0 $&  &  &&  \\
$\Xi^-$&  &  &  &  \\
\hline
\backslashbox{$B'$}{$B$} & \multicolumn{2}{c}{$\Xi^0$} & \multicolumn{2}{c}{$\Xi^-$}\\
\hline
$p$& & & &   \\
$n$&  &  &  &  \\
$\Lambda$&  &  &  &  \\
$\Sigma^0$&  &  &  &  \\
$\Sigma^+$& &  &  &  \\
$\Sigma^-$&  & &  &  \\
$\Xi^0 $&  \multicolumn{2}{c}{$-\frac{1}{3} \sqrt{2} \left(\sqrt{3} (D+3 F) \cos \epsilon+3 (F-D) \sin \epsilon\right)$} & & \\
$\Xi^-$&  &  &  \multicolumn{2}{c}{$-\frac{1}{3} \sqrt{2} \left(\sqrt{3} (D+3 F) \cos \epsilon+3 (D-F) \sin \epsilon\right)$} \\
\end{tabular}
\end{ruledtabular}
\end{centering}
\end{table}


\clearpage

\begin{table}[htbf]
\begin{centering}
\begin{ruledtabular}
\begin{tabular}{cccccc}
\multicolumn{6}{c}{$C_{BB\phi\phi'O_u}$}\\
\backslashbox{$B$}{$\phi\phi'$}&$\pi^0\pi^0$ & $\pi^+\pi^-$ & $K^0\overline{K}^0$ & $K^+K^-$ & $\eta \eta$ \\
\hline
$p$& & $\frac{1}{3} (\beta -2 \alpha )$ &  & $\frac{1}{6} (-5 \alpha -2 \beta )$ &  \\
$n$&  & $\frac{1}{3} (2 \alpha -\beta )$ &  & $\frac{1}{6} (-\alpha -4 \beta )$ &  \\
$\Lambda$&  &  &  & $\frac{1}{4} (\alpha -2 \beta )$ &  \\
$\Sigma^0$&  &  &  & $\frac{1}{4} (2 \beta -\alpha )$ &  \\
$\Sigma^+$&  & $\frac{1}{6} (-5 \alpha -2 \beta )$ &  & $\frac{1}{3} (\beta -2 \alpha )$ &  \\
$\Sigma^-$&  & $\frac{1}{6} (5 \alpha +2 \beta )$ &  & $\frac{1}{6} (\alpha +4 \beta )$ &  \\
$\Xi^0 $&  & $\frac{1}{6} (-\alpha -4 \beta )$ &  & $\frac{1}{3} (2 \alpha -\beta )$ &  \\
$\Xi^-$&  & $\frac{1}{6} (\alpha +4 \beta )$ &  & $\frac{1}{6} (5 \alpha +2 \beta )$ &  \\
\end{tabular}
\end{ruledtabular}
\end{centering}
\end{table}

\begin{table}[htbf]
\begin{centering}
\begin{ruledtabular}
\begin{tabular}{cccccc}
\multicolumn{6}{c}{$C_{BB\phi\phi'O_d}$}\\
\backslashbox{$B$}{$\phi\phi'$}&$\pi^0\pi^0$ & $\pi^+\pi^-$ & $K^0\overline{K}^0$ & $K^+K^-$ & $\eta \eta$ \\
\hline
$p$&  & $\frac{1}{3} (2 \alpha -\beta )$ & $\frac{1}{6} (-\alpha -4 \beta )$ &  &  \\
$n$&  & $\frac{1}{3} (\beta -2 \alpha )$ & $\frac{1}{6} (-5 \alpha -2 \beta )$ &  &  \\
$\Lambda$&  &  & $\frac{1}{4} (\alpha -2 \beta )$ &  &  \\
$\Sigma^0$&  &  & $\frac{1}{4} (2 \beta -\alpha )$ &  &  \\
$\Sigma^+$&  & $\frac{1}{6} (5 \alpha +2 \beta )$ & $\frac{1}{6} (\alpha +4 \beta )$ &  &  \\
$\Sigma^-$&  & $\frac{1}{6} (-5 \alpha -2 \beta )$ & $\frac{1}{3} (\beta -2 \alpha )$ &  &  \\
$\Xi^0 $&  & $\frac{1}{6} (\alpha +4 \beta )$ & $\frac{1}{6} (5 \alpha +2 \beta )$ &  &  \\
$\Xi^-$&  & $\frac{1}{6} (-\alpha -4 \beta )$ & $\frac{1}{3} (2 \alpha -\beta )$ &  &  \\
\end{tabular}
\end{ruledtabular}
\end{centering}
\end{table}

\begin{table}[htbf]
\begin{centering}
\begin{ruledtabular}
\begin{tabular}{cccccc}
\multicolumn{6}{c}{$C_{BB\phi\phi'O_s}$}\\
\backslashbox{$B$}{$\phi\phi'$}&$\pi^0\pi^0$ & $\pi^+\pi^-$ & $K^0\overline{K}^0$ & $K^+K^-$ & $\eta \eta$ \\
\hline
$p$&  &  & $\frac{1}{6} (\alpha +4 \beta )$ & $\frac{1}{6} (5 \alpha +2 \beta )$ &  \\
$n$&  &  & $\frac{1}{6} (5 \alpha +2 \beta )$ & $\frac{1}{6} (\alpha +4 \beta )$ &  \\
$\Lambda$&  &  & $\frac{1}{4} (2 \beta -\alpha )$ & $\frac{1}{4} (2 \beta -\alpha )$ &  \\
$\Sigma^0$&  &  & $\frac{1}{4} (\alpha -2 \beta )$ & $\frac{1}{4} (\alpha -2 \beta )$ &  \\
$\Sigma^+$&  &  & $\frac{1}{6} (-\alpha -4 \beta )$ & $\frac{1}{3} (2 \alpha -\beta )$ &  \\
$\Sigma^-$&  &  & $\frac{1}{3} (2 \alpha -\beta )$ & $\frac{1}{6} (-\alpha -4 \beta )$ &  \\
$\Xi^0 $&  &  & $\frac{1}{6} (-5 \alpha -2 \beta )$ & $\frac{1}{3} (\beta -2 \alpha )$ &  \\
$\Xi^-$&  &  & $\frac{1}{3} (\beta -2 \alpha )$ & $\frac{1}{6} (-5 \alpha -2 \beta )$ &  \\
\end{tabular}
\end{ruledtabular}
\end{centering}
\end{table}

\begin{table}[htbf]
\begin{centering}
\begin{ruledtabular}
\begin{tabular}{ccccccccc}
\multicolumn{9}{c}{$C_{BT\pi^0}C^{-1}$}\\
\backslashbox{$T$}{$B$}&$p$ & $n$ & $\Lambda$ & $\Sigma^0$ & $\Sigma^+$ & $\Sigma^-$ & $\Xi^0$ & $\Xi^-$ \\
\hline
$\Delta^{++}$&  &  &  &  &  &  &  &  \\
$\Delta^+$& $\sqrt{\frac{2}{3}}  \cos \epsilon$ &  &  &  &  &  &  &  \\
$\Delta^0$&  & $\sqrt{\frac{2}{3}}  \cos \epsilon$ &  &  &  &  &  &  \\
$\Delta^-$&  &  &  &  &  &  &  &  \\
$\Sigma^{*0}$&  &  & $-\frac{ \cos \epsilon}{\sqrt{2}}$ & $\frac{ \sin \epsilon}{\sqrt{2}}$ &  &  &  &  \\
$\Sigma^{*+}$&  &  &  &  & $-\frac{ \left(\cos \epsilon+\sqrt{3} \sin \epsilon\right)}{\sqrt{6}}$ &  &  &  \\
$\Sigma^{*-}$&  &  &  &  &  & $\frac{ \left(\sqrt{3} \sin \epsilon-\cos \epsilon\right)}{\sqrt{6}}$ &  &  \\
$\Xi^{*0}$&  &  &  &  &  &  & $-\frac{ \left(\cos \epsilon+\sqrt{3} \sin \epsilon\right)}{\sqrt{6}}$ &  \\
$\Xi^{*-}$&  &  &  &  &  &  &  & $\frac{ \left(\sqrt{3} \sin \epsilon-\cos \epsilon\right)}{\sqrt{6}}$ \\
$\Omega^-$&  &  &  &  &  &  &  &  \\
\end{tabular}
\end{ruledtabular}
\end{centering}
\end{table}

\begin{table}[htbf]
\begin{centering}
\begin{ruledtabular}
\begin{tabular}{ccccccccc}
\multicolumn{9}{c}{$C_{BT\pi^+}C^{-1}$}\\
\backslashbox{$T$}{$B$}&$p$ & $n$ & $\Lambda$ & $\Sigma^0$ & $\Sigma^+$ & $\Sigma^-$ & $\Xi^0$ & $\Xi^-$ \\
\hline
$\Delta^{++}$&  &  &  &  &  &  &  &  \\
$\Delta^+$&  &  &  &  &  &  &  &  \\
$\Delta^0$& $\frac{1}{\sqrt{3}}$ &  &  &  &  &  &  &  \\
$\Delta^-$&  & 1 &  &  &  &  &  &  \\
$\Sigma^{*0}$&  &  &  &  & $-\frac{1}{\sqrt{6}}$ &  &  &  \\
$\Sigma^{*+}$&  &  &  &  &  &  &  &  \\
$\Sigma^{*-}$&  &  & $-\frac{1}{\sqrt{2}}$ & $\frac{1}{\sqrt{6}}$ &  &  &  &  \\
$\Xi^{*0}$&  &  &  &  &  &  &  &  \\
$\Xi^{*-}$&  &  &  &  &  &  & $-\frac{1}{\sqrt{3}}$ &  \\
$\Omega^-$&  &  &  &  &  &  &  &  \\
\end{tabular}
\end{ruledtabular}
\end{centering}
\end{table}

\begin{table}[htbf]
\begin{centering}
\begin{ruledtabular}
\begin{tabular}{ccccccccc}
\multicolumn{9}{c}{$C_{BT\pi^-}C^{-1}$}\\
\backslashbox{$T$}{$B$}&$p$ & $n$ & $\Lambda$ & $\Sigma^0$ & $\Sigma^+$ & $\Sigma^-$ & $\Xi^0$ & $\Xi^-$ \\
\hline
$\Delta^{++}$& $-1$ &  &  &  &  &  &  &  \\
$\Delta^+$&  & $-\frac{1}{\sqrt{3}}$ &  &  &  &  &  &  \\
$\Delta^0$&  &  &  &  &  &  &  &  \\
$\Delta^-$&  &  &  &  &  &  &  &  \\
$\Sigma^{*0}$&  &  &  &  &  & $\frac{1}{\sqrt{6}}$ &  &  \\
$\Sigma^{*+}$&  &  & $\frac{1}{\sqrt{2}}$ & $\frac{1}{\sqrt{6}}$ &  &  &  &  \\
$\Sigma^{*-}$&  &  &  &  &  &  &  &  \\
$\Xi^{*0}$&  &  &  &  &  &  &  & $\frac{1}{\sqrt{3}}$ \\
$\Xi^{*-}$&  &  &  &  &  &  &  &  \\
$\Omega^-$&  &  &  &  &  &  &  &  \\
\end{tabular}
\end{ruledtabular}
\end{centering}
\end{table}

\begin{table}[htbf]
\begin{centering}
\begin{ruledtabular}
\begin{tabular}{ccccccccc}
\multicolumn{9}{c}{$C_{BTK^0}C^{-1}$}\\
\backslashbox{$T$}{$B$}&$p$ & $n$ & $\Lambda$ & $\Sigma^0$ & $\Sigma^+$ & $\Sigma^-$ & $\Xi^0$ & $\Xi^-$ \\
\hline
$\Delta^{++}$&  &  &  &  &  &  &  &  \\
$\Delta^+$&  &  &  &  &  &  &  &  \\
$\Delta^0$&  &  &  &  &  &  &  &  \\
$\Delta^-$&  &  &  &  &  &  &  &  \\
$\Sigma^{*0}$&  & $-\frac{1}{\sqrt{6}}$ &  &  &  &  &  &  \\
$\Sigma^{*+}$& $ -\frac{1}{\sqrt{3}}$  &  &  &  &  &  &  \\
$\Sigma^{*-}$&  &  &  &  &  &  &  &  \\
$\Xi^{*0}$&  &  & $\frac{1}{\sqrt{2}}$ & $\frac{1}{\sqrt{6}}$ &  &  &   &  \\
$\Xi^{*-}$&  &  &  &  &  & $\frac{1}{\sqrt{3}}$ &  &  \\
$\Omega^-$&  &  &  &  &  &  &  & $1$ \\
\end{tabular}
\end{ruledtabular}
\end{centering}
\end{table}

\begin{table}[htbf]
\begin{centering}
\begin{ruledtabular}
\begin{tabular}{ccccccccc}
\multicolumn{9}{c}{$C_{BTK^+}C^{-1}$}\\
\backslashbox{$T$}{$B$}&$p$ & $n$ & $\Lambda$ & $\Sigma^0$ & $\Sigma^+$ & $\Sigma^-$ & $\Xi^0$ & $\Xi^-$ \\
\hline
$\Delta^{++}$&  &  &  &  &  &  &  &  \\
$\Delta^+$&  &  &  &  &  &  &  &  \\
$\Delta^0$&  &  &  &  &  &  &  &  \\
$\Delta^-$&  &  &  &  &  &  &  &  \\
$\Sigma^{*0}$& $ \frac{1}{\sqrt{6}}$ &  &  &  &  &  &  &  \\
$\Sigma^{*+}$&   &  &  &  &  &  &  \\
$\Sigma^{*-}$&  & $\frac{1}{\sqrt{3}}$  &  &  &  &  &  &  \\
$\Xi^{*0}$&  &  &  & & $-\frac{1}{\sqrt{3}}$ &  &   &  \\
$\Xi^{*-}$&  &  & $-\frac{1}{\sqrt{2}}$ & $\frac{1}{\sqrt{6}}$  &  &  &  &  \\
$\Omega^-$&  &  &  &  &  &  & $-1$  & \\
\end{tabular}
\end{ruledtabular}
\end{centering}
\end{table}

\begin{table}[htbf]
\begin{centering}
\begin{ruledtabular}
\begin{tabular}{ccccccccc}
\multicolumn{9}{c}{$C_{BTK^-}C^{-1}$}\\
\backslashbox{$T$}{$B$}&$p$ & $n$ & $\Lambda$ & $\Sigma^0$ & $\Sigma^+$ & $\Sigma^-$ & $\Xi^0$ & $\Xi^-$ \\
\hline
$\Delta^{++}$&  &  &  &  & 1 &  &  &  \\
$\Delta^+$&  &  &  & $ -\sqrt{\frac{2}{3}}$  &  &  &  &  \\
$\Delta^0$&  &  &  &  &  & $-\frac{1}{\sqrt{3}}$ &  &  \\
$\Delta^-$&  &  &  &  &  &  &  &  \\
$\Sigma^{*0}$&  &  &  &  &  &  &  & $-\frac{1}{\sqrt{6}}$ \\
$\Sigma^{*+}$&   &  &  &  &  & $ \frac{1}{\sqrt{3}}$ &  \\
$\Sigma^{*-}$&  &   &  &  &  &  &  &  \\
$\Xi^{*0}$&  &  &  & &  &  &   &  \\
$\Xi^{*-}$&  &  &  &  &  &  &  &  \\
$\Omega^-$&  &  &  &  &  &  &   & \\
\end{tabular}
\end{ruledtabular}
\end{centering}
\end{table}

\begin{table}[htbf]
\begin{centering}
\begin{ruledtabular}
\begin{tabular}{ccccccccc}
\multicolumn{9}{c}{$C_{BT\overline{K}^0}C^{-1}$}\\
\backslashbox{$T$}{$B$}&$p$ & $n$ & $\Lambda$ & $\Sigma^0$ & $\Sigma^+$ & $\Sigma^-$ & $\Xi^0$ & $\Xi^-$ \\
\hline
$\Delta^{++}$&  &  &  &  &  &  &  &  \\
$\Delta^+$&  &  &  &   & $\frac{1}{\sqrt{3}}$ &  &  &  \\
$\Delta^0$&  &  &  & $-\sqrt{\frac{2}{3}}$ &  &  &  &  \\
$\Delta^-$&  &  &  &  &  & $-1$  &  &  \\
$\Sigma^{*0}$&  &  &  &  &  &  & $\frac{1}{\sqrt{6}}$ &  \\
$\Sigma^{*+}$&   &  &  &  &  &  &  \\
$\Sigma^{*-}$&  &   &  &  &  &  &  & $-\frac{1}{\sqrt{3}}$ \\
$\Xi^{*0}$&  &  &  & &  &  &   &  \\
$\Xi^{*-}$&  &  &  &  &  &  &  &  \\
$\Omega^-$&  &  &  &  &  &  &   & \\
\end{tabular}
\end{ruledtabular}
\end{centering}
\end{table}

\begin{table}[htbf]
\begin{centering}
\begin{ruledtabular}
\begin{tabular}{ccccccccc}
\multicolumn{9}{c}{$C_{BT\eta}C^{-1}$}\\
\backslashbox{$T$}{$B$}&$p$ & $n$ & $\Lambda$ & $\Sigma^0$ & $\Sigma^+$ & $\Sigma^-$ & $\Xi^0$ & $\Xi^-$ \\
\hline
$\Delta^{++}$&  &  &  &  &  &  &  &  \\
$\Delta^+$& $ -\sqrt{\frac{2}{3}} \sin \epsilon$ &  &  &   & &  &  &  \\
$\Delta^0$&  & $-\sqrt{\frac{2}{3}} \sin \epsilon$ &  & &  &  &  &  \\
$\Delta^-$&  &  &  &  &  &   &  &  \\
$\Sigma^{*0}$&  &  & $\frac{\sin \epsilon}{\sqrt{2}}$ & $\frac{\cos \epsilon}{\sqrt{2}}$  &  &  &  &  \\
$\Sigma^{*+}$&   &  &  & $\frac{\sin \epsilon-\sqrt{3} \cos \epsilon}{\sqrt{6}}$ &  &  &  \\
$\Sigma^{*-}$&  &   &  &  &  & $\frac{\sqrt{3} \cos \epsilon+\sin \epsilon}{\sqrt{6}}$  &  & \\
$\Xi^{*0}$&  &  &  & &  &  & $\frac{\sin \epsilon-\sqrt{3} \cos \epsilon}{\sqrt{6}}$  &  \\
$\Xi^{*-}$&  &  &  &  &  &  &  & $\frac{\sqrt{3} \cos \epsilon+\sin \epsilon}{\sqrt{6}}$ \\
$\Omega^-$&  &  &  &  &  &  &   & \\
\end{tabular}
\end{ruledtabular}
\end{centering}
\end{table}

\begin{table}[htbf]
\begin{centering}
\begin{ruledtabular}
\begin{tabular}{ccccccccccc}
\multicolumn{11}{c}{$C_{T'TO_u}(\gamma - \frac{\gamma'}{3})^{-1}$}\\
\backslashbox{$T'$}{$T$}& $\Delta^{++}$ & $\Delta^{+}$ & $\Delta^{0}$ & $\Delta^{-}$ & $\Sigma^{*0}$ & $\Sigma^{*+}$ & $\Sigma^{*-}$ & $\Xi^{*0}$ & $\Xi^{*-}$ & $\Omega^-$ \\
\hline
$\Delta^{++}$& 1 &  &  &  &  &  &  & & & \\
$\Delta^+$&  & $\frac{2}{3}$ &  &   & &  &  & & & \\
$\Delta^0$&  &  & $\frac{1}{3}$ & &  &  &  & & & \\
$\Delta^-$&  &  &  &  &  &   &  & & & \\
$\Sigma^{*0}$&  &  &   &  & $\frac{1}{3}$ &  & & & & \\
$\Sigma^{*+}$&   &  &  & &  & $\frac{2}{3}$ &  & & &\\
$\Sigma^{*-}$&  &   &  &  &  &   &  & & & \\
$\Xi^{*0}$&  &  &  & &  &  &  & $\frac{1}{3}$ & & \\
$\Xi^{*-}$&  &  &  &  &  &  &  & & & \\
$\Omega^-$&  &  &  &  &  &  &   & & &\\
\end{tabular}
\end{ruledtabular}
\end{centering}
\end{table}

\begin{table}[htbf]
\begin{centering}
\begin{ruledtabular}
\begin{tabular}{ccccccccccc}
\multicolumn{11}{c}{$C_{T'TO_d}(\gamma - \frac{\gamma'}{3})^{-1}$}\\
\backslashbox{$T'$}{$T$}& $\Delta^{++}$ & $\Delta^{+}$ & $\Delta^{0}$ & $\Delta^{-}$ & $\Sigma^{*0}$ & $\Sigma^{*+}$ & $\Sigma^{*-}$ & $\Xi^{*0}$ & $\Xi^{*-}$ & $\Omega^-$ \\
\hline
$\Delta^{++}$&  &  &  &  &  &  &  & & & \\
$\Delta^+$&  & $\frac{1}{3}$ &  &   & &  &  & & & \\
$\Delta^0$&  &  & $\frac{2}{3}$ & &  &  &  & & & \\
$\Delta^-$&  &  &  &  1 &  &   &  & & & \\
$\Sigma^{*0}$&  &  &   &  & $\frac{1}{3}$ &  & & & & \\
$\Sigma^{*+}$&   &  &  & &  &  &  & & &\\
$\Sigma^{*-}$&  &   &  &  &  &   & $\frac{2}{3}$ & & & \\
$\Xi^{*0}$&  &  &  & &  &  &  & & & \\
$\Xi^{*-}$&  &  &  &  &  &  &  & & $\frac{1}{3}$ & \\
$\Omega^-$&  &  &  &  &  &  &   & & &\\
\end{tabular}
\end{ruledtabular}
\end{centering}
\end{table}

\begin{table}[htbf]
\begin{centering}
\begin{ruledtabular}
\begin{tabular}{ccccccccccc}
\multicolumn{11}{c}{$C_{T'TO_s}(\gamma - \frac{\gamma'}{3})^{-1}$}\\
\backslashbox{$T'$}{$T$}& $\Delta^{++}$ & $\Delta^{+}$ & $\Delta^{0}$ & $\Delta^{-}$ & $\Sigma^{*0}$ & $\Sigma^{*+}$ & $\Sigma^{*-}$ & $\Xi^{*0}$ & $\Xi^{*-}$ & $\Omega^-$ \\
\hline
$\Delta^{++}$&  &  &  &  &  &  &  & & & \\
$\Delta^+$&  &  &  &   & &  &  & & & \\
$\Delta^0$&  &  &  & &  &  &  & & & \\
$\Delta^-$&  &  &  &   &  &   &  & & & \\
$\Sigma^{*0}$&  &  &   &  & $\frac{1}{3}$ &  & & & & \\
$\Sigma^{*+}$&   &  &  & &  & $\frac{1}{3}$ &  & & &\\
$\Sigma^{*-}$&  &   &  &  &  &   & $\frac{1}{3}$ & & & \\
$\Xi^{*0}$&  &  &  & &  &  &  & $\frac{2}{3}$ & & \\
$\Xi^{*-}$&  &  &  &  &  &  &  & & $\frac{2}{3}$ & \\
$\Omega^-$&  &  &  &  &  &  &   & & & 1\\
\end{tabular}
\end{ruledtabular}
\end{centering}
\end{table}

\begin{table}[htbf]
\begin{centering}
\begin{ruledtabular}
\begin{tabular}{ccccccccc}
\multicolumn{9}{c}{$C_{BTO_{\Delta u}}\omega^{-1}$}\\
\backslashbox{$T$}{$B$}&$p$ & $n$ & $\Lambda$ & $\Sigma^0$ & $\Sigma^+$ & $\Sigma^-$ & $\Xi^0$ & $\Xi^-$ \\
\hline
$\Delta^{++}$&  &  &  &  &  &  &  &  \\
$\Delta^+$& $\frac{1}{\sqrt{3}}$ &  &  &   & &  &  &  \\
$\Delta^0$&  & $\frac{1}{\sqrt{3}}$ &  & &  &  &  &  \\
$\Delta^-$&  &  &  &  &  &   &  &  \\
$\Sigma^{*0}$&  &  & $ -\frac{1}{2}$ & $\frac{1}{2 \sqrt{3}}$   &  &  &  &  \\
$\Sigma^{*+}$&   &  &  & & $ -\frac{1}{\sqrt{3}}$  &  &  &  \\
$\Sigma^{*-}$&  &   &  &  &  &  &  & \\
$\Xi^{*0}$&  &  &  & &  &  & $-\frac{1}{\sqrt{3}}$ &  \\
$\Xi^{*-}$&  &  &  &  &  &  &  &\\
$\Omega^-$&  &  &  &  &  &  &   & \\
\end{tabular}
\end{ruledtabular}
\end{centering}
\end{table}

\begin{table}[htbf]
\begin{centering}
\begin{ruledtabular}
\begin{tabular}{ccccccccc}
\multicolumn{9}{c}{$C_{BTO_{\Delta d}}\omega^{-1}$}\\
\backslashbox{$T$}{$B$}&$p$ & $n$ & $\Lambda$ & $\Sigma^0$ & $\Sigma^+$ & $\Sigma^-$ & $\Xi^0$ & $\Xi^-$ \\
\hline
$\Delta^{++}$&  &  &  &  &  &  &  &  \\
$\Delta^+$& $ -\frac{1}{\sqrt{3}}$ &  &  &   & &  &  &  \\
$\Delta^0$&  & $-\frac{1}{\sqrt{3}}$  &  & &  &  &  &  \\
$\Delta^-$&  &  &  &  &  &   &  &  \\
$\Sigma^{*0}$&  &  &  $\frac{1}{2}$ & $\frac{1}{2 \sqrt{3}}$  &  &  &  &  \\
$\Sigma^{*+}$&   &  &  &  &  &  & & \\
$\Sigma^{*-}$&  &   &  &  &  & $\frac{1}{\sqrt{3}}$ &  & \\
$\Xi^{*0}$&  &  &  & &  &  & &  \\
$\Xi^{*-}$&  &  &  &  &  &  &  & $ \frac{1}{\sqrt{3}}$ \\
$\Omega^-$&  &  &  &  &  &  &   & \\
\end{tabular}
\end{ruledtabular}
\end{centering}
\end{table}

\begin{table}[htbf]
\begin{centering}
\begin{ruledtabular}
\begin{tabular}{ccccccccc}
\multicolumn{9}{c}{$C_{BTO_{\Delta s}}\omega^{-1}$}\\
\backslashbox{$T$}{$B$}&$p$ & $n$ & $\Lambda$ & $\Sigma^0$ & $\Sigma^+$ & $\Sigma^-$ & $\Xi^0$ & $\Xi^-$ \\
\hline
$\Delta^{++}$&  &  &  &  &  &  &  &  \\
$\Delta^+$&  &  &  &   & &  &  &  \\
$\Delta^0$&  &  &  & &  &  &  &  \\
$\Delta^-$&  &  &  &  &  &   &  &  \\
$\Sigma^{*0}$&  &  &  & $-\frac{1}{\sqrt{3}}$  &  &  &  &  \\
$\Sigma^{*+}$&   &  &  &  & $ \frac{1}{\sqrt{3}}$  &  & & \\
$\Sigma^{*-}$&  &   &  &  &  & $-\frac{1}{\sqrt{3}}$  &  & \\
$\Xi^{*0}$&  &  &  & &  &  & $\frac{1}{\sqrt{3}}$ &  \\
$\Xi^{*-}$&  &  &  &  &  &  &  & $-\frac{1}{\sqrt{3}}$ \\
$\Omega^-$&  &  &  &  &  &  &   & \\
\end{tabular}
\end{ruledtabular}
\end{centering}
\end{table}

\end{widetext}

%
\clearpage


\end{document}